\documentclass{article}

\usepackage[utf8]{inputenc}
\usepackage[T1]{fontenc}

\usepackage{amsmath}
\usepackage{amsthm}
\usepackage{amsfonts}
\usepackage{amssymb}
\usepackage[mathic=true]{mathtools}

\usepackage{booktabs}
\usepackage{multirow}
\usepackage{graphicx}
\usepackage{thmtools}
\usepackage{mleftright}
\usepackage{subcaption}
\usepackage{xcolor}
\usepackage{xspace}
\usepackage{nicefrac}
\usepackage{microtype}
\usepackage{url}
\usepackage{paralist}
\usepackage{enumitem}

\usepackage[english,linesnumbered,vlined,ruled,nokwfunc,algo2e]{algorithm2e}
\DontPrintSemicolon
\SetKwComment{note}{$\triangleright$ }{}
\SetFuncSty{textsc}
\SetCommentSty{textit}
\SetDataSty{texttt}
\SetKwInput{Initial}{Initial}
\SetKwInput{Parameter}{Param.}
\SetKwInput{Input}{Input}
\SetKwInput{Data}{Data}
\SetKwInput{Output}{Output}
\ResetInOut{Result}
\let\oldnl\nl%
\newcommand{\nonl}{\renewcommand{\nl}{\let\nl\oldnl}}%
\SetKw{KwAnd}{and}
\SetKw{KwOr}{or}
\SetKw{KwXor}{xor}
\SetKw{KwNot}{not}
\SetKw{Parallel}{parallel}
\SetKw{Return}{return}
\SetKw{Break}{break}

\usepackage[final,nonatbib]{neurips_2025}
\usepackage{siunitx}
\usepackage[hidelinks]{hyperref}
\usepackage{cleveref}

\newtheorem{theorem}{Theorem}[section]
\newtheorem{lemma}[theorem]{Lemma}
\newtheorem{definition}[theorem]{Definition}
\newtheorem{observation}[theorem]{Observation}
\newtheorem{corollary}[theorem]{Corollary}

\crefname{theorem}{Theorem}{Theorems}
\Crefname{theorem}{Theorem}{Theorems}
\crefname{lemma}{Lemma}{Lemmas}
\Crefname{lemma}{Lemma}{Lemmas}
\crefname{definition}{Definition}{Definitions}
\Crefname{definition}{Definition}{Definitions}
\crefname{observation}{Observation}{Observations}
\Crefname{observation}{Observation}{Observations}
\crefname{corollary}{Corollary}{Corollaries}
\Crefname{corollary}{Corollary}{Corollaries}

\newcommand{\terminals}{\ensuremath{\mathcal{T}}}
\newcommand{\numgroups}{\ensuremath{|\mathcal{C}|}}
\newcommand{\budget}{\ensuremath{k}}

\newcommand{\qst}{\textsc{CoveringSteiner}\xspace}
\newcommand{\frtfull}{\textsc{FRT--Full}\xspace}
\newcommand{\frtbounded}{\textsc{FRT--Bounded}\xspace}
\newcommand{\sptfull}{\textsc{SPT--Full}\xspace}
\newcommand{\sptbounded}{\textsc{SPT--Bounded}\xspace}
\newcommand{\pdstrict}{\textsc{PD--Strict}\xspace}
\newcommand{\pdmatched}{\textsc{PD--Matched}\xspace}
\newcommand{\coveringsteiner}{\textsc{Covering Steiner}\xspace}
\newcommand{\algclosest}{\textsc{Closest}\xspace}
\newcommand{\algalternating}{\textsc{Alternating}\xspace}
\newcommand{\groupagnostic}{\textsc{Group-Agnostic}\xspace}

\newcommand{\aspect}{\Delta}

\title{Fair Minimum Labeling: Efficient Temporal Network Activations for Reachability and Equity}

\author{%
Lutz Oettershagen\\
University of Liverpool\\
\texttt{lutz.oettershagen@liverpool.ac.uk} \\
\And
Othon Michail\\
University of Patras\\
\texttt{othon.michail@upatras.gr} \\
}

\begin{document}

\maketitle

\begin{abstract}
Balancing resource efficiency and fairness is critical in networked systems that support modern learning applications. We introduce the \emph{Fair Minimum Labeling} (FML) problem: the task of designing a minimum-cost temporal edge activation plan that ensures each group of nodes in a network has sufficient access to a designated target set, according to specified coverage requirements. FML captures key trade-offs in systems where edge activations incur resource costs and equitable access is essential, such as distributed data collection, update dissemination in edge-cloud systems, and fair service restoration in critical infrastructure. We first give a structural characterisation of the single-terminal case, showing that it is equivalent to the rooted Covering Steiner problem. We prove that FML is NP-hard and admits no $((1-\epsilon)\ln \numgroups)$-approximation for $\numgroups$ groups, already on a star, while for any fixed number of groups it inherits a constant-factor approximation and remains APX-hard. We then present probabilistic approximation algorithms for the two-group, single-terminal case: an algorithm whose tree subroutine is exact, hence optimal on tree-structured networks and $\mathcal{O}(\log |V|)$ in expectation on general graphs, together with a faster bicriteria variant whose coverage violation degrades gracefully with the merge depth of the tree computation. For practical scalability, we additionally introduce a graph-native variant based on a shortest-path-tree reduction. 
Empirical results show that FML enforces group-level fairness with substantially lower activation cost than the baselines.
\end{abstract}

\section{Introduction}\label{sec:intro}

Resource efficiency and fairness represent dual challenges in designing modern machine learning and networked systems. From large-scale federated learning managing communication budgets across heterogeneous devices~\cite{kairouz2021advances}, to recommender systems scheduling updates and notifications to diverse user groups~\cite{jalali2023fairness}, and distributed sensor networks operating under stringent energy constraints~\cite{huaizhou2013fairness}, the underlying theme is constant: interactions are dynamic, resources are limited, and equity across participants or data sources is paramount~\cite{caton2024fairness,dong2023fairness,mao2024green}. In such systems, the graph structure captures potential interactions, but the temporal dimension, i.e., when connections are active, governs the actual flow of information, updates, and influence.

Optimizing these temporal interactions solely for aggregate performance or minimal static infrastructure cost often yields suboptimal or unfair outcomes. Simple heuristics might overload communication bottlenecks, drain energy reserves rapidly, or systematically neglect harder-to-reach nodes or groups representing minority populations, specific event types, or distinct data distributions. This can lead to biased data collection pipelines, skewed model training, feedback loops that amplify inequity, and inefficient use of limited resources like energy or bandwidth.

Therefore, principled mechanisms are needed that explicitly co-optimize efficiency and fairness in these dynamic network settings. We consider systems where a central controller has the ability to decide when communication links are active. Our focus is on designing temporal activation patterns over an underlying static graph $G=(V,E)$ that satisfy strong temporal reachability guarantees for distinct node groups, while minimizing the total number of activations. Nodes may represent users, sensors, or agents belonging to specific populations ($V_1,\ldots, V_{\numgroups}$), and fairness dictates that each group must maintain adequate connectivity over time to designated target nodes. Minimizing the number of edge activations directly translates to reduced energy consumption, communication overhead, and computational load, which is central for sustainable system design~\cite{mao2024green,schwartz2020green}.

To address this, we introduce the \textbf{Fair Minimum Labeling (FML)} problem framework. Conceptually, FML seeks to determine \emph{which} edges in the graph need to be activated and \emph{when}, such that the required number of nodes from \emph{each specified group} can reach the target nodes via temporal paths formed by these time-ordered activations. The primary objective is to achieve this guaranteed fair reachability using the {minimum total number of edge activations over time}. FML thus provides a formal basis for optimizing resource sparsity while enforcing equitable temporal access.

The introduction of FML as a novel optimization problem is motivated by both practical relevance and theoretical gaps. First, it applies to a wide range of systems where temporal resource allocation directly affects group equity. In distributed data gathering for learning applications, such as those involving environmental sensors, mobile agents, or geographically diverse data streams~\cite{nassef2022survey,verbraeken2020survey,vatter2023evolution}, FML enables minimal-cost activation plans that ensure all groups are fairly represented in the collected data. In edge-cloud systems, it helps decide when communication links should be active so that all device groups can receive timely model updates under bandwidth constraints. Similarly, in infrastructure restoration, FML can guide activation sequences that ensure fair reactivation of services. Second, FML fills a critical gap in the literature. Existing fairness-aware graph methods, such as fair clustering~\cite{kleindessner2019guarantees}, node embeddings~\cite{rahman2019fairwalk}, and influence maximization~\cite{tsang2019group,windham2024fast}, are primarily designed for static settings and do not model temporal interactions or activation cost. Fairness has been studied in wireless systems~\cite{huaizhou2013fairness}, and temporal connectivity has been explored in dynamic networks~\cite{kempe2000connectivity,mertzios2013temporal}, but fairness constraints remain absent in algorithmic temporal graph design. FML addresses this gap by explicitly minimizing temporal edge activations, treated as a proxy for resource cost, while enforcing group-level temporal reachability. In contrast to decentralized approaches such as classical federated learning~\cite{kairouz2021advances}, where participation is uncoordinated, FML assumes centralized control over link activations, allowing the enforcement of fairness guarantees over time. This design objective distinguishes FML from prior work focused on static structures or continuous flows.

\textbf{Our main contributions can be summarized as follows:}
\begin{itemize}
	\item We introduce the \emph{Fair Minimum Labeling (FML)} problem, a new framework for designing temporally sparse interaction patterns that guarantee group-wise temporal reachability under global resource constraints, and we show that it generalizes classical temporal connectivity problems.

	\item We prove that FML with a single terminal is NP-complete and admits no $((1-\epsilon)\ln\numgroups)$-approximation in the number of groups $\numgroups$ under Feige's standard complexity assumption for Set Cover, already on a star (\Cref{theorem:maincomplexity}) and already when every node belongs to at most one group (\Cref{theorem:disjointcolors}). For a constant number of groups the problem is APX-hard but inherits a $2\numgroups$-approximation (\Cref{theorem:twocolor,theorem:constant}).

	\item We develop a tree-embedding framework for the case of two groups reaching a single terminal, yielding two algorithms: (i) one whose tree subroutine is exact, so that it is optimal on tree-structured inputs and $\mathcal{O}(\log |V|)$ in expectation on general graphs, and (ii) a faster bicriteria variant that guarantees $\mathcal{O}(\log |V|)$ expected cost and a coverage-relaxation factor of $(1+\varepsilon)^{D}$, where $D$ is the \emph{merge depth} of the embedding tree (\Cref{def:mergedepth}).

	\item We introduce a graph-native variant (\Cref{sec:spt_fml}), which replaces the randomized FRT hierarchy and graph-projection phase by a shortest-path-tree reduction, safely prunes the tree to its colored core, and provides a bounded-relaxation parameterization of the same tree dynamic program. 

	\item We empirically evaluate our algorithms both in a multi-source learning task and in scalability study on the Pokec social network~\cite{takac2012data}. 
\end{itemize}

The omitted proofs are provided in \Cref{appendix}.

\textbf{Revision note.} This manuscript is a corrected and revised version of the originally published paper~\cite{oettershagen2026fair}. The theoretical results and their analysis have been revised, and the presentation and experimental evaluation have been updated and improved accordingly.

\section{Preliminaries}\label{sec:prelim}

We denote the set of positive integers (natural numbers without zero) by $\mathbb{N} = \{1, 2, 3, \ldots\}$. For $\ell \in \mathbb{N}$, we use $[\ell]$ to denote the set $\{1, 2, \ldots, \ell\}$. The powerset of a set $X$ is denoted by $\mathcal{P}(X)$.

A \emph{static graph} is an ordered pair $G=(V, E)$, where $V$ is a finite set of \emph{nodes} (or vertices) and $E \subseteq \{\{u,v\} \mid u, v \in V, u \neq v\}$ is a finite set of undirected \emph{edges}. We assume static graphs are simple (no self-loops, no multiple edges between the same pair of nodes).
A \emph{colored static graph} $G=(V, E, c)$ with color set $\mathcal{C}=[c_{\max}]$ has an additional function $c:V\rightarrow \mathcal{P}(\mathcal{C})$ assigning a (possibly empty) set of colors $c(u)$ to each node $u\in V$; a node $u$ with $c(u)=\emptyset$ is called \emph{uncolored}.
For $c\in\mathcal{C}$ we write $V_c = \{v \in V \mid c \in c(v)\}$.
Let $n = |V|$ and $m = |{E}|$. We assume $G$ to be connected.

A \emph{temporal graph} is an ordered pair $\mathcal{G}=(V, \mathcal{E})$, where $V$ is a finite set of nodes and $\mathcal{E}$ is a finite set of \emph{temporal edges}. Each temporal edge is a pair $e=(\{u,v\}, \tau)$, where $\{u,v\}$ is an undirected pair of distinct nodes from $V$, and $\tau \in \mathbb{N}$ is the \emph{timestamp} indicating when the interaction between $u$ and $v$ occurs.

Given a static graph $G=(V,E)$, a \emph{temporal labeling} is a function $\lambda: E \rightarrow \mathcal{P}(\mathbb{N})$ that assigns a set of timestamps (possibly empty) to each static edge $e \in E$.
A labeling $\lambda$ induces a temporal graph $G_\lambda = (V, E_\lambda)$, where the node set is the same as the static graph's, and the set of temporal edges is defined as
$E_\lambda = \{ (\{u,v\}, \tau) \mid e=\{u,v\} \in E \text{ and } \tau \in \lambda(e) \}$.
The \emph{size} (or \emph{cost}) of a labeling $\lambda$, denoted by $|\lambda|$, is the total number of assigned timestamps, which corresponds to the number of temporal edges in the induced temporal graph, that is,
$|\lambda| = |E_\lambda| = \sum_{e \in E} |\lambda(e)|$.
The \emph{support} of $\lambda$ is the set of edges that receive at least one timestamp, $\mathrm{supp}(\lambda)=\{e\in E \mid \lambda(e)\neq\emptyset\}$; note that $|\mathrm{supp}(\lambda)|\le|\lambda|$.

A \emph{temporal path} (or \emph{time-respecting path}) in a temporal graph $\mathcal{G}=(V, \mathcal{E})$ from node $u$ to node $v$ is a sequence of temporal edges $( \{v_0, v_1\}, \tau_1 ), ( \{v_1, v_2\}, \tau_2 ), \ldots, ( \{v_{p-1}, v_p\}, \tau_p )$ from $\mathcal{E}$ such that $v_0 = u$, $v_p = v$, and the timestamps are strictly increasing, i.e., $\tau_1 < \tau_2 < \dots < \tau_p$.
A node $v$ is \emph{temporally reachable} from a node $u$ in $\mathcal{G}$ if such a temporal path exists from $u$ to $v$.

We define the \emph{reachability indicator function} $r(u,v)$ with respect to a given temporal graph $\mathcal{G}=(V, \mathcal{E})$. For any two nodes $u, v \in V$:
\[
r(u,v) =
\begin{cases}
	1 & \text{if } v \text{ is temporally reachable from } u \text{ in } \mathcal{G}, \\
	0 & \text{otherwise.}
\end{cases}
\]
By convention, we consider every node to be temporally reachable from itself (via a path of length zero), so $r(v,v) = 1$ for all $v \in V$. The specific temporal graph $\mathcal{G}$ (often an induced graph $G_\lambda$) for which reachability is considered will always be clear from the context.
Finally, for $U, W\subseteq V$, let $r(U,W)=\sum_{u\in U, v\in W} r(u,v)$ be the sum of pairwise reachability of nodes in $U$ to nodes in $W$.

\begin{observation}\label{obs:normalize}
	Let $\lambda$ be a temporal labeling of $G$ with $|\lambda|=k$. There is a temporal labeling $\lambda'$ with $|\lambda'|=k$, with all timestamps in $[k]$, and with exactly the same temporal reachability relation as~$\lambda$.
\end{observation}
\begin{proof}
	Temporal reachability depends only on the relative order of the timestamps. Replacing the $i$-th smallest distinct timestamp used by $\lambda$ with the value $i$ preserves the order and uses at most $k$ distinct values.
\end{proof}

\section{Related Work}\label{sec:related}

The Fair Minimum Labeling (FML) problem intersects several research areas, primarily temporal graph algorithms, network optimization with fairness considerations, and resource-efficient AI.

\textbf{Temporal Graph Connectivity and Labeling.}
A core component of FML is finding a minimum-cost set of temporal edge activations. Mertzios et al.~\cite{mertzios2013temporal} (and its journal version~\cite{mertzios2019temporal}) introduced the study of temporal network design in multi-labeled temporal graphs, where the goal is to optimize some global cost measure of the graph labeling, such as the total number of timestamps, while satisfying specific temporal reachability properties. The \emph{Minimum Labeling (ML)} problem is such a temporal network design problem, seeking the minimum number of labels to make an entire static graph temporally connected. For its directed version on strongly connected digraphs,~\cite{mertzios2013temporal} had provided bounds on the number of labels needed, proving that 2 labels per arc and at most $4(|V|-1)$ total labels are always sufficient. Using similar constructions on undirected graphs, Akrida et al.~\cite{akrida2017complexity} proved that $2|V|-3$ total labels are sufficient, later refined by Klobas et al.~\cite{klobas2024complexity} to $2|V|-4$ under certain conditions, who also showed ML is polynomially solvable.
While ML aims for global temporal connectivity, FML focuses on targeted reachability for specific node groups to a designated terminal set, under explicit fairness (coverage) constraints for each group, and a cost minimization objective. Klobas et al.~\cite{klobas2024complexity} also introduced extensions like \emph{Minimum Steiner Labeling (MSL)}, where the goal is to pairwise temporally connect a predefined set of terminal nodes. Our FML generalizes ML and MSL.
Other research has explored concepts like temporally-connected spanning subgraphs~\cite{axiotis2016size}, temporal core decompositions~\cite{galimberti2020span,oettershagen2025edge}, and various notions of temporal paths and reachability~\cite{kempe2000connectivity,casteigts2012time,michail2016introduction,gionis2024mining,oettershagen2023higher,oettershagen2022temporal,mutzel2019enumeration}. However, these works generally do not incorporate group fairness constraints or the specific cost model of minimizing temporal edge activations as FML does.

\textbf{Fairness in Graph Mining and Network Optimization.}
There is a significant body of work on incorporating fairness into static graph algorithms and network optimization, driven by the need for equitable outcomes in various applications.
This includes, e.g., fair clustering~\cite{ahmadian2019clustering,chierichetti2017fair,kleindessner2019guarantees}, fair node embeddings~\cite{rahman2019fairwalk,bose2019compositional}, fair influence maximization~\cite{tsang2019group,windham2024fast}, and graph learning tasks~\cite{chen2024fairness,dai2024comprehensive,navarro2024fair,zhou2024fairness,oettershagen2024finding}.
While these contributions are crucial for advancing fairness in graph-based AI, they predominantly operate on static graphs and often define fairness differently (e.g., proportionality in output, balanced representation in a static structure). For example, a related problem in static graphs is the \emph{Balanced Connected Subgraph (BCS)} problem~\cite{bhore2022balanced} which seeks a connected subgraph with balanced representation from different node colors. FML, in contrast, addresses fairness through \emph{temporal path-based accessibility} for predefined groups, directly optimizing the dynamic activation of interactions over time to meet these fairness goals while minimizing resource usage.
Moreover, in networked systems, fairness is a long-standing concern in resource allocation. This includes fair bandwidth sharing in communication networks~\cite{kelly1997charging}, fair scheduling in wireless systems~\cite{huaizhou2013fairness,vaidya2000distributed}, and equitable service provisioning. In edge-cloud environments, fair scheduling of updates (e.g., for recommender systems~\cite{jalali2023fairness}) or computation offloading is critical. FML contributes to this line of work by providing a formal model for achieving group-level temporal reachability fairness specifically through the lens of minimizing discrete edge activations, a key proxy for communication cost, energy, or computational overhead, aligning with sustainability principles~\cite{mao2024green,schwartz2020green}. Unlike other problems that focus on flow rates or latency targets, FML provides hard guarantees on temporal path existence for specified quotas from diverse groups.

\textbf{Connection to Combinatorial Optimization.}
By \Cref{lemma:steiner}, the single-terminal case of FML is exactly the rooted \emph{Covering Steiner} problem: find a minimum-cost tree spanning at least a prescribed number of vertices from every group. This problem was introduced by Konjevod and Ravi and studied by Konjevod, Ravi and Srinivasan~\cite{konjevod2002covering}; Gupta and Srinivasan~\cite{gupta2006covering} give the currently best guarantee, $\mathcal{O}(\log N\log \numgroups)$ on tree instances and $\mathcal{O}(\log n\log N\log \numgroups)$ on general graphs, where $n$ is the number of vertices, $N$ the maximum group size and $\numgroups$ the number of groups. They also observe an $\mathcal{O}(\numgroups)$-approximation obtained by solving one rooted quota subproblem per group and taking the union, which is the argument behind our \Cref{theorem:constant}. Covering Steiner in turn generalizes two classical problems. With all requirements equal to one it is the group Steiner tree problem, which is $\mathcal{O}(\log^{2}n\log \numgroups)$-approximable on general graphs~\cite{garg2000polylogarithmic} and $\Omega(\log^{2-\epsilon}\numgroups)$-hard~\cite{halperin2003polylogarithmic}. With a single group it is $k$-MST, which admits a $2$-approximation~\cite{garg2005saving}, and with a single group and full coverage it is the classical Steiner tree problem, APX-hard~\cite{chlebik2008steiner} and approximable within $1.39$~\cite{byrka2013steiner}.

Our results sit inside this landscape as follows. Our lower bound (\Cref{theorem:maincomplexity}) is obtained by a self-contained, approximation-preserving reduction from Set Cover~\cite{karp1972,feige1998threshold}; it is weaker than the bound inherited from~\cite{halperin2003polylogarithmic}, but it is elementary, it holds on a star, and it makes the role of the labeling cost explicit. Both lower bounds require an unbounded number of groups. In the opposite direction, the same equivalence hands us an upper bound: for any fixed number of groups, single-terminal FML inherits a constant-factor approximation on general graphs (\Cref{theorem:constant}), which is better than the $\mathcal{O}(\log n)$ factor our own tree-embedding algorithm achieves. We therefore position the algorithms of \Cref{sec:approximation} on what they do that the generic machinery does not: the tree subroutine is exact, hence optimal on tree-structured inputs, and the bicriteria variant trades coverage for speed in a controlled way. The temporal dimension of FML becomes genuinely binding when $\rho\ge2$, i.e., when a counted node must reach multiple terminals; this is precisely the regime not covered by \Cref{lemma:steiner}.

\section{The fair minimum labeling problem}\label{sec:fml}

\begin{figure}
	\centering
	\begin{subfigure}[b]{0.31\textwidth}
		\centering
		\includegraphics[width=0.8\linewidth]{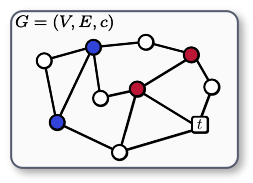}
		\caption{Example of a colored static graph $G$ with one terminal node $t\in V$ shown as square. }
		\label{fig:example1a}
	\end{subfigure}\hfill%
	\begin{subfigure}[b]{0.31\textwidth}
		\centering
		\includegraphics[width=0.8\linewidth]{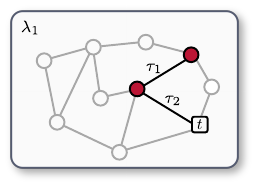}
		\caption{Temporal labeling $\lambda_1$ connects $50\%$ of the colored nodes to $t$, but only red nodes.}
		\label{fig:example1b}
	\end{subfigure}\hfill%
		\begin{subfigure}[b]{0.31\textwidth}
		\centering
		\includegraphics[width=0.8\linewidth]{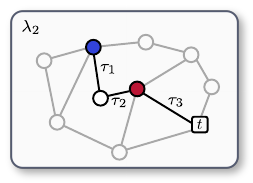}
		\caption{Temporal labeling $\lambda_2$ connects $50\%$ of the colored nodes in equal proportion to $t$.}
		\label{fig:example1c}
	\end{subfigure}
	\caption{A simple toy example showing a non-equitable and an equitable labeling of a small colored graph. The labelings $\lambda_1$ and $\lambda_2$ assign time stamps $\tau_1<\tau_2<\tau_3$ to edges. Both connect 50\% of the colored nodes, but only $\lambda_2$ connects both colors resulting in higher costs of three. Therefore, under a color requirement of $50\%$ for each color, only $\lambda_2$ is a valid solution for the FML problem.
	}
	\label{fig:example}
\end{figure}

We now formalize the central optimization problem studied in this work: designing temporally sparse activation patterns over a static graph to guarantee group-level access to designated target nodes under strict resource constraints. In many learning-integrated systems, such as data aggregation, distributed monitoring, or update dissemination, ensuring both efficiency and equitable access over time is essential. Existing frameworks, however, lack mechanisms to enforce fairness through temporal reachability while minimizing total activation effort.

The \emph{Fair Minimum Labeling (FML)} problem addresses this gap by integrating three key aspects: (i) cost-aware edge activation planning, (ii) per-group temporal reachability guarantees, and (iii) flexible, group-specific coverage requirements. The goal is to compute a temporal edge labeling that minimizes total activation cost while satisfying fairness constraints across all groups (see \Cref{app:example} for an application example).

\begin{definition}[Fair Minimum Labeling (FML)]\label{def:fml}
	Given a colored static graph $G = (V, E, c)$ with color set $\mathcal{C}$, a set of terminals $\terminals \subseteq V$, a parameter $\rho\in [\,|\terminals|\,]$, and a requirement $q_c \in \mathbb{N}_0$ for each color $c \in \mathcal{C}$, the goal is to find a temporal labeling $\lambda : E \rightarrow \mathcal{P}(\mathbb{N})$ minimizing the total number of assigned time labels
	$
	|\lambda| = \sum_{e \in E} |\lambda(e)|
	$
	subject to
	\[
		\bigl|\{\, v \in V_c \;\mid\; \textstyle\sum_{t \in \terminals} r(v,t) \ \ge\ \rho \,\}\bigr| \ \ge\ q_c
		\qquad \text{for every } c \in \mathcal{C},
	\]
	where reachability is with respect to the temporal graph $G_\lambda$ induced by $\lambda$. That is, for every color $c$, at least $q_c$ nodes of $V_c$ must temporally reach at least $\rho$ terminals.
\end{definition}

A fractional specification is obtained by setting $q_c = \lceil \alpha_c |V_c| \rceil$ for a fraction $\alpha_c\in[0,1]$. We use this form throughout \Cref{sec:approximation}. Note that $\rho=1$ is the case in which every counted node must reach \emph{some} terminal, and $\rho=|\terminals|$ the case in which it must reach \emph{all} of them.

\Cref{fig:example} shows a simple example.
In the corresponding decision version, we are additionally given a budget $\budget \in \mathbb{N}$ and asked whether there exists a temporal labeling $\lambda$ such that $|\lambda| \leq \budget$ while satisfying the coverage constraints for all colors. We also note that a symmetric variant of the problem can be defined by requiring temporal reachability \emph{from} (rather than \emph{towards}) the terminal set $\terminals$.

The FML problem generalizes both the Minimum Labeling (ML) and the Minimum Steiner Labeling (MSL) problems (see \Cref{obs:generalization} in the appendix). Since MSL is NP-hard~\cite{klobas2024complexity}, FML is NP-hard as well.

\subsection{A structural characterisation of the single-terminal case}\label{sec:steiner}

Before turning to complexity, we record a structural fact about the case $\terminals=\{t\}$. It says that for a single terminal the temporal dimension of FML collapses: an optimal solution never needs more than one timestamp per edge, and the problem reduces to a purely static covering problem. We use it for both the lower bounds of \Cref{sec:hardness} and the correctness of the dynamic program of \Cref{sec:exact_tree}.

\begin{definition}[Rooted Covering Steiner~\cite{konjevod2002covering}]\label{def:qst}
	Given a colored static graph $G=(V,E,c)$, a root $t\in V$ and requirements $q_c\in\mathbb{N}_0$ for $c\in\mathcal{C}$, the \qst problem asks for a subtree $S$ of $G$ with $t\in V(S)$ minimizing $|E(S)|$ subject to $|V_c\cap V(S)|\ge q_c$ for all $c\in\mathcal{C}$. It generalizes the group Steiner tree problem (all $q_c=1$) and the $k$-MST problem (a single group containing all vertices).
\end{definition}

\begin{lemma}\label{lemma:steiner}
	Let $I$ be an instance of FML with $\terminals=\{t\}$, and let $I'$ be the \qst instance with the same graph, root $t$ and the same requirements $q_c$. Then $\mathrm{OPT}_{\mathrm{FML}}(I) = \mathrm{OPT}_{\qst}(I')$. Moreover, an optimal FML solution can be taken to assign exactly one timestamp to each edge of an optimal subtree and no timestamp to any other edge, and it can be computed from the subtree in linear time.
\end{lemma}

\begin{proof}
	($\ge$) Let $\lambda$ be feasible for $I$ and let $W\subseteq V$ be the set of nodes that temporally reach $t$ under $\lambda$. Every $v\in W$ is connected to $t$ by a path in the static graph $(V,\mathrm{supp}(\lambda))$, since a temporal path is in particular a walk. Hence $W\cup\{t\}$ lies in the connected component $K$ of $t$ in $(V,\mathrm{supp}(\lambda))$. A spanning tree $S$ of $K$ satisfies $t\in V(S)$, $W \subseteq V(S)$ and $|E(S)| \le |\mathrm{supp}(\lambda)| \le |\lambda|$, and it meets every requirement because $\lambda$ does. Thus $\mathrm{OPT}_{\qst}(I')\le \mathrm{OPT}_{\mathrm{FML}}(I)$.

	($\le$) Let $S$ be feasible for $I'$, root it at $t$, and let $h$ be its height. Assign to the edge whose lower endpoint has depth $d$ the single timestamp $h-d+1$, and assign no timestamp elsewhere; this yields a labeling $\lambda_S$ with $|\lambda_S| = |E(S)|$. Along the path from any $v \in V(S)$ to $t$ the depth strictly decreases, so the timestamps strictly increase, and the path is time-respecting. Hence every node of $S$ reaches $t$ and $\lambda_S$ is feasible for $I$, giving $\mathrm{OPT}_{\mathrm{FML}}(I) \le \mathrm{OPT}_{\qst}(I')$.
\end{proof}

We write $\mathrm{FML}_w$ for the variant in which each edge carries an activation cost $w(e)>0$ and the objective is $\sum_{e\in E} w(e)\,|\lambda(e)|$; FML is the case $w\equiv1$. \Cref{lemma:steiner} and its proof hold verbatim for $\mathrm{FML}_w$ with $|E(S)|$ replaced by $\sum_{e\in E(S)}w(e)$.
	\Cref{lemma:steiner} is specific to a single target. With $|\terminals|\ge2$ and $\rho\ge2$ a node must reach several terminals along time-respecting paths that share edges, and a single timestamp per edge no longer suffices; this is where FML departs from static Steiner problems and where the results of Klobas et al.~\cite{klobas2024complexity} on Steiner labelings become relevant. The algorithms of \Cref{sec:approximation} address the single-terminal case, and we discuss the multi-terminal extension in \Cref{sec:conclusion}.

\subsection{Complexity}\label{sec:hardness}

\begin{theorem}\label{theorem:maincomplexity}
	The decision version of FML is NP-complete, already for a single terminal and for instances in which every requirement is $q_c=1$. Moreover, for every fixed $\epsilon>0$, unless $\mathrm{NP}\subseteq\mathrm{DTIME}[N^{\mathcal{O}(\log\log N)}]$, where $N$ denotes the universe size of the Set Cover instances used in the reduction, no polynomial-time algorithm approximates single-terminal FML within a factor $(1-\epsilon)\ln\numgroups$, where $\mathcal{C}$ is the set of color classes. This holds already when $G$ is a star.
\end{theorem}

The proof is in \Cref{appendix}. The construction is an approximation-preserving reduction from Set Cover on a star: the leaves are the sets, a leaf carries one color per element it contains, and activating a leaf edge is the same as choosing the set. That it holds on a star is worth noting, since it means the hardness is a matter of the number of groups alone and not of the graph structure. If one insists that every node carry at most one color the following variant applies instead.

\begin{theorem}\label{theorem:disjointcolors}
	The NP-completeness and $(1-\epsilon)\ln\numgroups$-inapproximability conclusions of \Cref{theorem:maincomplexity} continue to hold when restricted to instances in which every node carries at most one color.
\end{theorem}

Its proof, also in \Cref{appendix}, replaces the star by a construction in which every set-element incidence gets its own degree-one node, so that no colored node can act as a relay for another, and each set node is joined to the terminal by a private chain whose length forces the budget accounting.

\begin{theorem}\label{theorem:constant}
	Single-terminal FML with $\numgroups$ color classes admits a polynomial-time $2\numgroups$-approximation. In particular, the two-group case admits a $4$-approximation on general graphs.
\end{theorem}
\begin{proof}
	By \Cref{lemma:steiner} it suffices to approximate \qst. For each color $c$, apply the $2$-approximation of Garg~\cite{garg2005saving} to the rooted $q_c$-MST instance on group $V_c$, obtaining a tree $S_c$ that contains $t$ and at least $q_c$ vertices of $V_c$, with $|E(S_c)| \le 2\,\mathrm{OPT}_c$. An optimal \qst solution is feasible for every such single-group instance, so $\mathrm{OPT}_c \le \mathrm{OPT}$ for every $c$. The union $S=\bigcup_c S_c$ contains $t$, is connected because every $S_c$ contains $t$, and satisfies every requirement. Hence any spanning tree of $S$ has at most
	\[
	\sum_{c\in\mathcal{C}} |E(S_c)| \le 2\numgroups\,\mathrm{OPT}
	\]
	edges. Applying the ($\le$) direction of \Cref{lemma:steiner} yields a temporal labeling of the same size. This is the $\mathcal{O}(\numgroups)$ union-of-groups argument of Gupta and Srinivasan~\cite{gupta2006covering}, transported through \Cref{lemma:steiner}.
\end{proof}

\begin{theorem}\label{theorem:twocolor}
	FML with a single terminal and a single color class with $q_{c_1}=|V_{c_1}|$ is NP-hard and APX-hard; the same holds a fortiori for two color classes. Together with \Cref{theorem:constant}, this places the constant-group case squarely between a constant lower bound and a constant upper bound, so no super-constant inapproximability bound can hold there.
\end{theorem}
\begin{proof}
	By \Cref{lemma:steiner} this case is exactly the Steiner tree problem with terminal set $V_{c_1}\cup\{t\}$ and unit edge costs, which is NP-hard and APX-hard~\cite{chlebik2008steiner}; conversely it is approximable within $1.39$~\cite{byrka2013steiner}. Adding a second color class with $q_{c_2}=0$ does not change the instance.
\end{proof}

\section{Approximation Algorithms}\label{sec:approximation}

Given the hardness results in \Cref{sec:hardness}, we now present approximation algorithms for FML instances with a single terminal.
We focus on the case where each node is blue, red, or uncolored, i.e., $\mathcal{C} = \{B, R\}$ and $c(v)\in\{\{B\},\{R\},\emptyset\}$ for every $v$.
For convenience, we use the set notation $B \subseteq V$ and $R \subseteq V$ to directly refer to the nodes of each group.
Furthermore, we set the requirements to $q_B = \lceil\alpha |B|\rceil$ and $q_R = \lceil\alpha |R|\rceil$ for a fixed parameter $\alpha \in [0,1]$, indicating that an $\alpha$-fraction of both $B$ and $R$ must reach the terminal.
This two-group setting captures a common fairness use case and provides a natural foundation for future extensions to multiple groups or terminals.

\subsection{FRT-based reduction}\label{sec:frt_reduction}

Our approach leverages the framework of \emph{probabilistic tree embeddings}, a powerful technique for approximating arbitrary metric spaces with simpler tree metrics.
The concept was introduced by Bartal~\cite{bartal1996probabilistic}, who showed that any finite metric space can be probabilistically approximated by a distribution over tree metrics, incurring only a logarithmic distortion in expectation. This was later improved by Fakcharoenphol, Rao, and Talwar (FRT)~\cite{fakcharoenphol2003tight}, who established that any $n$-point metric space can be embedded into a distribution over dominating tree metrics with an expected stretch of $\mathcal{O}(\log n)$, and that this bound is tight.

\begin{definition}\label{def:embedding}
Let $(X, d)$ be a finite metric space. A \emph{probabilistic tree embedding} is a randomized mapping from $(X, d)$ into a distribution $\mathcal{D}$ over weighted trees $T$ whose leaf set is $X$, such that for all $x, y \in X$:
\begin{enumerate}
	\item (non-contraction) for every $T$ in the support of $\mathcal{D}$, $d(x,y) \leq d_T(x,y)$;
	\item (expected stretch) $\mathbb{E}_{T \sim \mathcal{D}}[d_T(x,y)] \leq \mathcal{O}(\log |X|) \cdot d(x,y)$.
\end{enumerate}
\end{definition}

	The tree $T$ produced by FRT on $(V,d_G)$ is a $2$-hierarchically well-separated tree: its leaves are the vertices $V$, its internal nodes are clusters of a laminar family, and all edges between level $i$ and level $i-1$ have weight $\Theta(2^{i})$, so that edge weights decay geometrically along any root-to-leaf path. Its height is $H = \mathcal{O}(\log \aspect)$, where $\aspect = \max_{u\neq v} d_G(u,v) / \min_{u\neq v}d_G(u,v)$ is the aspect ratio of the metric. For the unweighted shortest-path metric of a connected graph $\aspect \le n-1$ and hence $H=\mathcal{O}(\log n)$. After contracting internal nodes with a single child, which changes no leaf-to-leaf distance and can only decrease the height, $T$ has at most $2n-1$ nodes. For a node $x$ of $T$ we fix an arbitrary leaf $\mathrm{rep}(x)$ in the subtree of $x$, with $\mathrm{rep}(v)=v$ for a leaf $v$. Colors are carried by the leaves only; internal nodes are uncolored.

\begin{algorithm2e}[htb]
	\caption{Approximation Framework for FML with Single Terminal}
	\label[algorithm]{alg:approx1}
	\Input{Graph $G=(V,E)$, $B\subseteq V$, $R\subseteq V$, terminal $t\in V$, $\alpha \in [0,1]$, and $\varepsilon>0$ for the bicriteria variant}
	\Output{Temporal graph labeling $\lambda_G$}

	Compute the shortest path metric $d_G$ on $G$.\;
	Sample a tree $T$ from the FRT distribution $\mathcal{D}$ for $(V, d_G)$~\cite{fakcharoenphol2003tight}; contract unary internal nodes and fix representatives.\;
	Root the tree $T$ at the leaf $t$.\;
	Using the exact dynamic program, let $S$ be a minimum-weight subtree with quotas $q_B,q_R$. Using the bicriteria dynamic program, let $S$ be the subtree represented by a minimum-cost retained root label satisfying $b\ge \lceil q_B/\xi\rceil$ and $r\ge \lceil q_R/\xi\rceil$, where $\xi=(1+\varepsilon)^D$.\;
	Project $S$ to a connected subgraph of $G$ and label it (\Cref{sec:mapback}), obtaining $\lambda_G$.\;

	\Return $\lambda_{G}$
\end{algorithm2e}

\Cref{alg:approx1} outlines our framework. We use the shortest-path metric $d_G$, where $d_G(u,v)$ is the minimum number of edges on a $u$-$v$ path in $G$. By the weighted extension of \Cref{lemma:steiner}, solving $\mathrm{FML}_w$ on $T$ with root $t$ is the same as finding a minimum-weight subtree $S$ of $T$ that contains $t$ and enough colored vertices; we therefore state the tree subroutines in that form and write
\[ C_T(S) = \sum_{e \in E(S)} w(e) \]
for its weighted cost. The exact subroutine (\Cref{sec:exact_tree}) minimizes $C_T$ subject to the constraints; the bicriteria subroutine (\Cref{sec:approx_tree}) returns a subtree of cost at most the exact optimum whose color counts may fall short of the requirements by a factor $\xi$.

\subsection{An exact algorithm on weighted trees}\label{sec:exact_tree}

We compute labels at the nodes of $T$ bottom-up, starting from the leaves.
Each node of the input tree may be blue, red, or uncolored; in an FRT tree, only nodes corresponding to original graph vertices are colored. We use labels $(b,r,c)$, where $b$ and $r$ count the selected blue and red vertices and $c$ is the total selected edge weight. The local contribution of a node $v$ is $([\mathrm{col}(v)=B],[\mathrm{col}(v)=R],0)$, as in \Cref{alg:bottomup-tree-labels}.

Let $v$ be an inner node with $\ell$ children $u_1,\ldots,u_\ell$ carrying label sets $L_1,\dots,L_\ell$. A solution at $v$ picks at most one label per child, which determines the (possibly empty) selected part of that child's subtree, so before pruning there are at most
\[ \prod_{j=1}^{\ell} \bigl(1+|L_j|\bigr) \]
combinations at $v$. At each node we keep at most one label $(b,r,c)$ per pair $(b,r)$, namely one of minimum cost $c$. Since $b\le|B|\le n$ and $r\le|R|\le n$, there are $\mathcal{O}(n^2)$ pairs, which also bounds the number of retained labels per node.
\Cref{alg:bottomup-tree-labels} in the appendix gives the pseudocode.

\begin{theorem}\label{theorem:runningtimetree}
	The weighted variant $\mathrm{FML}_w$ on a tree with $\mathcal{O}(n)$ nodes, a single terminal and two color classes can be solved exactly in $\mathcal{O}(n^{5})$ time and $\mathcal{O}(n^{3})$ space.
\end{theorem}

\subsection{A bicriteria approximation on weighted trees}\label{sec:approx_tree}

The $\mathcal{O}(n^5)$ running time of the exact algorithm limits its scalability.
The bottleneck is the quadratic number of labels per node, which makes a pairwise merge cost $\mathcal{O}(n^4)$. The idea is therefore to reduce the number of labels in a principled way by grid rounding: group labels with similar $(b,r)$ values into a common bucket and keep only the cheapest label per bucket.

\begin{definition}\label{def:buckets}
	For $\varepsilon>0$ let $\mathrm{Bucket}(0) = 0$ and, for an integer $x\ge1$, $\mathrm{Bucket}(x) = 1+\lfloor \log_{1+\varepsilon} x\rfloor$, so that bucket $0$ contains only the value $0$, bucket $1$ contains the values in $[1,1+\varepsilon)$, and bucket $i\ge1$ contains the values in $[(1+\varepsilon)^{i-1},(1+\varepsilon)^{i})$. Two values in the same non-zero bucket differ by a factor of less than $1+\varepsilon$.
\end{definition}

Since $b\le n$ and $r\le n$, there are $\beta = \mathcal{O}(\log_{1+\varepsilon} n)^2 = \mathcal{O}(\varepsilon^{-2}\log^{2}n)$ bucket pairs for $\varepsilon\in(0,1]$, and after rounding each label set has at most $\beta$ entries.

Where the rounding is applied matters. Rounding only after the full label set of a node has been assembled would give one rounding per tree level, but it requires materialising $\prod_j(1+|L_j|)$ intermediate labels. We therefore round after \emph{every} pairwise merge, which keeps every intermediate set of size at most $\beta$, and we account for the error accordingly.

\begin{definition}\label{def:mergedepth}
	For every inner node $v$ with children $u_1,\ldots,u_{\ell_v}$, merge the child contributions according to a balanced binary tree. For a child $u$, its contribution set is
	\[
		L^+_{v,u}
		=
		\{(0,0,0)\}
		\cup
		\{(b,r,c+w(v,u)):(b,r,c)\in L_u\},
	\]
	where $(0,0,0)$ represents not selecting the child's subtree. The color contribution of $v$ itself is added exactly after these merges and is not bucketed.

	The \emph{merge depth} $D$ is the maximum number of bucket-pruned pairwise merges encountered along any leaf-to-root computation path. Hence
	\[
		D\le
		\sum_{v\in P}\left\lceil\log_2\max\{1,\ell_v\}\right\rceil
		\le H\lceil\log_2 n\rceil
		=\mathcal{O}(H\log n),
	\]
	where $P$ is a heaviest root-to-leaf path of $T$. For the FRT tree of an unweighted graph metric, $D=\mathcal{O}(\log^{2} n)$.
\end{definition}

The algorithm processes the local merge gadgets bottom-up: the leaves of each local merge gadget carry the exact contribution sets $L^+_{v,u}$, each merge node combines the two label sets of its children by adding counts and costs, and the resulting set is immediately reduced to at most $\beta$ labels by keeping, for each bucket pair, one label of minimum cost. The color contribution of the current vertex is then added exactly.

\begin{theorem}\label{theorem:weighted_approx}
	Let $\varepsilon\in(0,1]$, let $T$ be the input tree with root $t$ and merge depth $D$, and let $\xi=(1+\varepsilon)^D$. Let $(b_{\mathrm{opt}},r_{\mathrm{opt}},c_{\mathrm{opt}})$ be a minimum-cost exact root label satisfying $b_{\mathrm{opt}}\ge q_B$ and $r_{\mathrm{opt}}\ge q_R$.

	At the root, let the bicriteria algorithm return a minimum-cost retained label $(b',r',c')$ satisfying
	\[
		b'\ge \widetilde q_B:=\left\lceil\frac{q_B}{\xi}\right\rceil,
		\qquad
		r'\ge \widetilde q_R:=\left\lceil\frac{q_R}{\xi}\right\rceil .
	\]
	Such a label exists and satisfies
	\begin{itemize}
		\item[(i)] $c'\le c_{\mathrm{opt}}$,
		\item[(ii)] $b'\ge \widetilde q_B\ge q_B/\xi$ and $r'\ge \widetilde q_R\ge q_R/\xi$.
	\end{itemize}
	It runs in $\mathcal{O}(n\varepsilon^{-4}\log^4 n)$ time and $\mathcal{O}(n\varepsilon^{-2}\log^2 n)$ space.
\end{theorem}

\begin{corollary}\label{cor:rescale}
	For a desired coverage relaxation $\hat\varepsilon\in(0,1]$, running the algorithm with $\varepsilon = \hat\varepsilon/(2\max\{1,D\})$ yields $\xi \le 1+\hat\varepsilon$, in time $\mathcal{O}(n\,\hat\varepsilon^{-4}\max\{1,D\}^{4}\log^{4}n)$, i.e., $\mathcal{O}(n\,\hat\varepsilon^{-4}\log^{12}n)$ for the FRT tree of an unweighted graph metric.
\end{corollary}

\subsection{Projection and approximation guarantee}\label{sec:frt_guarantee}\label{sec:mapback}

Let $S\subseteq T$ be the subtree returned by the tree subroutine. For every edge $\{x,y\}$ of $S$ we take a shortest path in $G$ between $\mathrm{rep}(x)$ and $\mathrm{rep}(y)$, and we let $\Pi(S)$ be the union of all these paths together with the vertices they traverse.

\begin{lemma}\label{lemma:projection}
	$\Pi(S)$ is a connected subgraph of $G$ containing $t$ and every colored leaf of $S$, and there is a constant $\kappa$ (depending only on the separation parameter of the FRT tree) with $|E(\Pi(S))| \le \kappa \cdot C_T(S)$. Consequently there is a temporal labeling $\lambda_G$ of $G$ with $|\lambda_G| \le \kappa\, C_T(S)$ under which every colored leaf of $S$ temporally reaches $t$, and $\lambda_G$ is computable in $\mathcal{O}(m + |E(\Pi(S))|)$ time.
\end{lemma}
\begin{proof}
	Connectivity is immediate: the paths are glued along the representatives, and $S$ is connected. For the cost, fix $\{x,y\}\in E(S)$ and orient it according to the original FRT hierarchy, before rerooting at $t$, so that $y$ is the ancestor of $x$. Since $\mathrm{rep}(x)$ and $\mathrm{rep}(y)$ are leaves in the subtree of $y$, non-contraction gives $d_G(\mathrm{rep}(x),\mathrm{rep}(y)) \le d_T(\mathrm{rep}(x),\mathrm{rep}(y))$, and by the geometric decay of edge weights along root-to-leaf paths in a $2$-HST the latter is at most $\kappa\,w(x,y)$ for an absolute constant $\kappa$. Summing over $E(S)$ gives the bound. The labeling then follows from the ($\le$) direction of \Cref{lemma:steiner} applied to a spanning tree of $\Pi(S)$ rooted at $t$: one timestamp per edge, assigned by decreasing depth.
\end{proof}

Note that the projection may additionally connect colored nodes that were not selected in $S$; this can only help, since it can only increase coverage.

\begin{theorem}\label{theorem:main}
	\Cref{alg:approx1} is a randomized algorithm for FML with one terminal and two colors on general graphs with the following guarantees, where $k^*_G$ denotes the optimal cost.
	\begin{itemize}
		\item With the exact tree algorithm (\Cref{sec:exact_tree}), the returned $\lambda_G$ satisfies the coverage constraints exactly and $\mathbb{E}[|\lambda_G|] = \mathcal{O}(\log n)\cdot k^*_G$; when $G$ is itself a tree the embedding step is unnecessary and the algorithm returns an \emph{optimal} solution. The running time is $\mathcal{O}(nm + n^{5})$ and the space is $\mathcal{O}(n^{3})$.
		\item With the bicriteria tree algorithm (\Cref{sec:approx_tree}), $\mathbb{E}[|\lambda_G|] = \mathcal{O}(\log n)\cdot k^*_G$ and the coverage requirements are relaxed by a factor $\xi=(1+\varepsilon)^{D}$, giving a randomized bicriteria $(\mathcal{O}(\log n),\xi)$-approximation. The running time is $\mathcal{O}(nm + n\,\varepsilon^{-4}\log^{4}n)$ and the space is $\mathcal{O}(n^2 + n\varepsilon^{-2}\log^{2}n)$.
	\end{itemize}
\end{theorem}

\subsection{A graph-native variant}\label{sec:spt_fml}

The FRT reduction above provides an expected-distortion guarantee, but constructing the metric hierarchy and projecting the selected FRT subtree back to the input graph introduce substantial preprocessing and reconstruction overhead. We therefore also consider a graph-native reduction that keeps the same rooted two-group tree dynamic program while replacing the FRT hierarchy by a shortest-path spanning tree (SPT). We call the resulting framework \emph{SPT-FML}. The input remains a general graph; the tree is only an internal reduction.

Let $G=(V,E)$ be connected with positive edge costs $c_e$ and terminal $t$. We construct a shortest-path spanning tree $T$ rooted at $t$, using BFS for unit-cost graphs and Dijkstra's algorithm for weighted graphs, with deterministic predecessor tie-breaking. Hence, for every $v\in V$,
\[
    d_T(t,v)=d_G(t,v).
\]
This equality is only a root-distance property: it does not imply that $d_T(u,v)=d_G(u,v)$ for arbitrary pairs $u,v$. In contrast to the FRT tree, the SPT contains only original graph vertices and original graph edges. Thus group membership remains attached directly to the original vertices, there are no artificial cluster vertices or representatives, and no projection back to $G$ is needed after the tree subproblem is solved. A selected SPT subtree can be labeled directly using the construction of \Cref{lemma:steiner}.

Before running the dynamic program, we repeatedly delete every leaf $v\notin B\cup R\cup\{t\}$ from $T$ and denote the remaining tree by $T_C$. This \emph{colored core} is a safe instance reduction for the chosen SPT: an uncolored leaf cannot lie on a path from a blue or red vertex to the root, so deleting it cannot remove any rooted subtree that contributes to either coverage quota. 
A colored internal vertex contributes its own color count
to every state at that vertex, as in \Cref{alg:bottomup-tree-labels};
this contribution is exact and introduces no additional bucketing or
merge-depth loss.

Both FRT-FML and SPT-FML use the bucketed two-dimensional dynamic program of \Cref{sec:approx_tree} on the resulting rooted tree. 
We distinguish the strict quotas
\[
q_B=\lceil\alpha |B|\rceil,
\qquad
q_R=\lceil\alpha |R|\rceil
\]
from the relaxed quotas used to select a root state. Let $D$ be the merge depth of the actual tree computation and let $\varepsilon>0$ be the bucketing parameter. With
\[
\xi=(1+\varepsilon)^D,
\]
we use
\[
\widetilde q_B=\left\lceil\frac{q_B}{\xi}\right\rceil,
\qquad
\widetilde q_R=\left\lceil\frac{q_R}{\xi}\right\rceil.
\]

For a desired end-to-end coverage relaxation $\hat\varepsilon\in(0,1]$, we choose
\[
\varepsilon
=\frac{\hat\varepsilon}{2\max\{1,D\}}.
\]
For $D\ge1$,
\[
\xi=(1+\varepsilon)^D
\le e^{\hat\varepsilon/2}
\le 1+\hat\varepsilon,
\]
and hence, for each $g\in\{B,R\}$,
\[
\widetilde q_g
\ge
\left\lceil\frac{q_g}{1+\hat\varepsilon}\right\rceil.
\]
Thus $\varepsilon$ controls the bucketing resolution, while $\hat\varepsilon$ specifies the desired end-to-end coverage relaxation.

To quantify the graph-to-tree reduction, we measure the maximum graph-edge stretch on the full SPT, before colored-core pruning,
\[
    s_T
    =\max_{\{u,v\}\in E}\frac{d_T(u,v)}{c_{uv}},
\]
and also record the mean graph-edge stretch. For the chosen SPT, the following instance-dependent comparison holds:
\[
    \operatorname{OPT}_T(q_B,q_R)
    \le s_T\,\operatorname{OPT}_G(q_B,q_R).
\]
Indeed, replace every edge of a feasible graph solution by the corresponding path in $T$; connectivity and covered vertices are preserved, while the total path cost is at most $s_T$ times the graph-solution cost. This is a diagnostic, not an FRT-style worst-case guarantee: an arbitrary shortest-path tree need not have $\mathcal{O}(\log n)$ stretch, and preservation of terminal-to-vertex distances does not preserve all pairwise distances. 

\subsection{Generalizations}\label{sec:generalizations}

The algorithms and analysis above are for the two-group (plus uncolored) case with a single terminal.
The framework generalizes to $\numgroups$ groups: the label becomes $(g_1,\ldots,g_{\numgroups},c)$, the count state space is $\mathcal{O}(n^{\numgroups})$, a pairwise merge takes $\mathcal{O}(n^{2\numgroups})$ time, and the exact dynamic program runs in $\mathcal{O}(n^{2\numgroups+1})$ time. This is polynomial for every fixed $\numgroups$ but not for growing $\numgroups$, which is consistent with \Cref{theorem:maincomplexity}: the instances witnessing the logarithmic lower bound have polynomially many groups, so no algorithm of this shape can be efficient there.

Non-uniform activation costs $w(e)$ can be supported by computing the shortest-path metric with respect to $w$ and embedding that metric. The $\mathcal{O}(\log n)$ guarantee on the expected weighted cost is unaffected, because \Cref{lemma:steiner} and \Cref{lemma:projection} are both cost-oblivious.
The bicriteria guarantee, however, degrades: by the FRT height bound above the height of the embedding is $\mathcal{O}(\log\aspect)$ for the aspect ratio $\aspect$ of the weighted metric, so $D = \mathcal{O}(\log \aspect\log n)$ and the coverage-relaxation factor becomes $(1+\varepsilon)^{\mathcal{O}(\log\aspect\log n)}$ unless $\varepsilon$ is rescaled as in \Cref{cor:rescale}.

Finally, in dynamic environments (e.g., node arrivals, failures or departures), partial recomputation of our labeling is feasible. The dynamic program can be rerun locally on affected subtrees, and minor changes in the graph structure often preserve the validity of the tree embedding. While full dynamic adaptation is beyond the scope of this work, the methods lend themselves to such extensions.

\section{Experiments}\label{sec:experiments}

We evaluate our algorithms in three complementary settings: a multi-source learning use case, a sensitivity study of the coverage-relaxation parameter, and a scalability study on the Pokec social network~\cite{takac2012data}. The multi-source experiment evaluates the effect of group-wise coverage on downstream learning. The sensitivity experiment studies how the relaxation parameter affects runtime, dynamic-programming state space, activation cost, and achieved coverage. The Pokec experiment evaluates the scalability of the practical methods on a large real-world network and compares the FRT and SPT reductions against the primal--dual baseline.

\textbf{Algorithms.}
We evaluate the algorithmic variants developed in the preceding sections
together with several baselines:
\begin{itemize}
	\item \frtfull uses the FRT framework of \Cref{alg:approx1}
	with the full tree dynamic program of \Cref{sec:exact_tree}.
	
	\item \frtbounded uses the same FRT framework with the
	bicriteria tree dynamic program of \Cref{sec:approx_tree} and bounded
	coverage relaxation.
	
	\item \sptfull and \sptbounded use the graph-native
	shortest-path-tree reduction of \Cref{sec:spt_fml} with the corresponding
	full and bounded tree dynamic programs.
	
	\item \pdstrict is a rooted primal--dual heuristic using the
	original group-wise coverage requirements, while \pdmatched
	uses the same relaxed requirements as the bounded method against which
	it is compared.
	
	\item \coveringsteiner is the certified two-group baseline
	corresponding to the constant-factor construction of
	\Cref{theorem:constant}.
	
	\item \algclosest, \algalternating, and
	\groupagnostic are used only in the multi-source learning
	experiment. \algclosest selects nearby sources until both groups
	are sufficiently covered, \algalternating alternates this selection
	between the groups, and \groupagnostic ignores group membership.
\end{itemize}

\subsection{Experimental setup}\label{sec:exp_setup}

All experiments contain two protected groups, denoted Blue and Red, and one
terminal. All graph edges have unit cost. Every returned connected solution
is reduced to a terminal-rooted tree and assigned a valid temporal labeling.
We therefore report activation cost as the number of selected tree edges,
equivalently the number of edge-time labels. Shared edges are counted only
once.

The full-DP variants and strict baselines are evaluated against the original
group-wise coverage targets. The bounded variants are bicriteria methods and
may intentionally receive smaller coverage
requirements. 
For the bounded variants, we set $\hat{\varepsilon}=0.10$.
Consequently, a relaxed method falling below the original
target is not an execution failure if it satisfies the requirement that it
was asked to solve. We compare solution costs directly only when methods
receive the same coverage requirements, or when the achieved coverage is
reported alongside the cost.

All comparisons use paired instances: methods evaluated in the same trial
receive the same graph, terminal, and protected groups. Reported runtimes are
end-to-end and include graph-to-tree preprocessing and, for the FRT variants,
projection back to the original graph.

We implemented all algorithms in Python 3.13. The experiments ran on a single machine with an Intel i5-1345U CPU and 32GB of main memory. The source code is available at \textcolor{black}{\url{https://gitlab.com/tgdesign/fml_code}}\,.

\subsection{Multi-source learning use case}\label{sec:multi_source}

We first study how group-wise coverage affects downstream learning.
Each trial uses a random geometric graph with 1,024 vertices and connection
radius $0.2$. The terminal is the graph vertex nearest the geometric corner
$(0,0)$. We select 64 Blue data sources closest to the terminal and 64 Red
data sources farthest from it. The original coverage requirement is 32
sources from each group.

Each covered source contributes eight two-dimensional training examples.
For Blue sources, the two Gaussian class centers are $(-2,1)$ and $(2,1)$;
for Red sources, they are $(1,2)$ and $(1,-2)$. The Gaussian standard
deviation is $0.55$. The construction gives the two groups different
classification rules, so representative training data from both groups is
needed for strong performance on both test populations.

For every trial, we train a shared classifier on the data collected from the
covered sources. The classifier uses two hidden layers of 32 ReLU units,
Adam optimization with learning rate $0.01$, batch size 64, and
$\ell_2$ regularization of $0.01$. Training uses at most 100 iterations, a
10\% validation split, and early stopping after ten iterations without
validation improvement. Independent balanced test sets contain 1,000
examples from each group.

\textbf{Results.}
\Cref{tab:scheduler_perf_std} reports the results. Runtime and cost are shown
as mean $\pm$ standard deviation over the ten paired trials; coverage,
accuracy, and the absolute difference between group accuracies are means.

\begin{table*}[t]
	\centering
	\caption{Multi-source learning results over ten paired trials.
		Methods are grouped by the coverage requirements they receive.
		Runtime and cost are mean $\pm$ standard deviation; the remaining
		entries are means.}
	\label{tab:scheduler_perf_std}
	
	\setlength{\tabcolsep}{5pt}
	\resizebox{\linewidth}{!}{
		\begin{tabular}{
				@{}
				l
				S[table-format=2.3] @{\,$\pm$\,} S[table-format=2.3]
				S[table-format=3.1] @{\,$\pm$\,} S[table-format=2.1]
				r
				r
				r
				@{}
			}
			\toprule
			\textbf{Method} &
			\multicolumn{2}{c}{\textbf{Runtime (s)}} &
			\multicolumn{2}{c}{\textbf{Cost}} &
			\textbf{Coverage B/R} &
			\textbf{Accuracy B/R} &
			\textbf{Acc.~gap} \\
			\midrule
			
			\multicolumn{8}{l}{\emph{Original coverage requirements}}\\
			\frtfull
			& 0.079 & 0.022
			& 75.3 & 4.6
			& $33.1/32.4$ & $.946/.954$ & $.030$ \\
			
			\sptfull
			& 0.136 & 0.021
			& 87.7 & 4.4
			& $32.0/32.0$ & $.935/.959$ & $.036$ \\
			
			\pdstrict
			& 0.724 & 0.021
			& 68.0 & 0.0
			& $32.0/32.0$ & $.950/.948$ & $.034$ \\
			
			\coveringsteiner
			& 47.505 & 28.288
			& 68.5 & 0.7
			& $32.3/32.0$ & $.939/.949$ & $.044$ \\
			
			\algclosest
			& 0.301 & 0.033
			& 118.2 & 12.8
			& $32.6/32.0$ & $.943/.948$ & $.048$ \\
			
			\algalternating
			& 0.322 & 0.035
			& 115.8 & 9.3
			& $32.1/32.0$ & $.940/.939$ & $.067$ \\
			
			\midrule
			\multicolumn{8}{l}{\emph{Bounded relaxation}}\\
			\frtbounded
			& 0.086 & 0.017
			& 74.1 & 4.2
			& $31.9/31.8$ & $.946/.953$ & $.031$ \\
			
			\sptbounded
			& 0.145 & 0.029
			& 84.9 & 4.1
			& $31.0/31.0$ & $.940/.957$ & $.027$ \\
			
			\pdmatched
			& 0.760 & 0.056
			& 66.0 & 0.0
			& $31.0/31.0$ & $.947/.951$ & $.035$ \\
			
			\midrule
			\multicolumn{8}{l}{\emph{Fairness-unaware baseline}}\\
			\groupagnostic
			& 0.191 & 0.030
			& 64.0 & 0.0
			& $64.0/0.0$ & $.996/.408$ & $.588$ \\
			
			\bottomrule
	\end{tabular}}
\end{table*}

Among methods satisfying the original group-wise requirements,
\pdstrict and \coveringsteiner have the lowest
mean costs, $68.0$ and $68.5$, followed by
\frtfull at $75.3$ and
\sptfull at $87.7$.
The two proximity-based heuristics are substantially more expensive, with
mean costs above~$115$.

The relaxed methods satisfy the coverage requirements they are given and
must be interpreted separately from the strict methods.
Under the bounded requirements,
\pdmatched has lower mean cost than
\sptbounded ($66.0$ versus $84.9$), but requires
$0.760$~s rather than $0.145$~s on average.

The clearest downstream effect appears for \groupagnostic.
Although its mean activation cost is only $64.0$, it selects all 64 Blue
sources and no Red source. Its resulting mean accuracies are $.996$ and
$.408$, giving an accuracy gap of $.588$.
Every group-aware method maintains high test accuracy for both groups.
Thus, low aggregate collection cost alone does not guarantee representative
data or balanced downstream performance.

\subsection{Effect of the relaxation parameter}
\label{sec:epsilon_sensitivity}

We isolate the effect of the bounded-relaxation parameter
$\hat{\varepsilon}$ on \frtbounded and \sptbounded using unweighted
Barab\'asi--Albert graphs with attachment parameter $3$.
The maximum-degree vertex is used as the terminal, and two disjoint,
equal-sized source groups are formed from vertices near to and far from the
terminal. The original requirement is to cover half of each group.
We consider
$\hat{\varepsilon}\in\{0.01,0.025,0.05,0.10,0.20,0.40\}$
over nine $(n,s)$ configurations,
$
(1024,64),\ (1024,128),\ (1024,256)$, 
$(2048,128),\ (2048,256),\ (2048,512)$, 
$(4096,128),\ (4096,256),\ (4096,512),
$
where $n$ is the number of vertices and $s$ the number of sources per group.
Each configuration uses ten paired random seeds. Within each paired comparison, the graph instance and tree construction are fixed and only $\hat{\varepsilon}$ is varied. All trials
completed successfully and satisfied the relaxed coverage requirement.

\begin{table*}[t]
	\centering
	\caption{Effect of $\hat{\varepsilon}$ on the bounded FRT and SPT
		algorithms. Entries are paired mean percentage reductions relative to
		$\hat{\varepsilon}=0.01$; larger values therefore indicate lower
		runtime, fewer retained DP states, lower activation cost, or lower
		minimum achieved group coverage. Runtime is solver time and States is
		the peak number of post-bucketing dynamic-programming states.}
	\label{tab:epsilon_sensitivity}
	\setlength{\tabcolsep}{4.5pt}
	\resizebox{0.8\linewidth}{!}{
		\begin{tabular}{@{}c rrrr rrrr@{}}
			\toprule
			& \multicolumn{4}{c}{\frtbounded}
			& \multicolumn{4}{c}{\sptbounded} \\
			\cmidrule(lr){2-5}\cmidrule(l){6-9}
			$\hat{\varepsilon}$
			& \textbf{Runtime} & \textbf{States} & \textbf{Cost} & \textbf{Coverage}
			& \textbf{Runtime} & \textbf{States} & \textbf{Cost} & \textbf{Coverage} \\
			\midrule
			$0.05$  & $2.2\%$ & $3.4\%$ & $1.8\%$ & $1.7\%$
			& $2.6\%$ & $3.3\%$ & $1.9\%$ & $1.7\%$ \\
			$0.10$  & $5.1\%$ & $8.6\%$ & $4.5\%$ & $4.5\%$
			& $6.7\%$ & $8.5\%$ & $5.0\%$ & $4.4\%$ \\
			$0.20$  & $9.7\%$ & $17.5\%$ & $9.7\%$ & $9.3\%$
			& $13.8\%$ & $17.4\%$ & $10.3\%$ & $9.3\%$ \\
			$0.40$  & $19.1\%$ & $31.8\%$ & $17.7\%$ & $17.5\%$
			& $27.9\%$ & $34.2\%$ & $19.1\%$ & $17.4\%$ \\
			\bottomrule
	\end{tabular}}
\end{table*}

\Cref{tab:epsilon_sensitivity} shows a clear effect of the relaxation
parameter. At $\hat{\varepsilon}=0.05$, the solver
time decreases by $2.2\%$ for \frtbounded and $2.6\%$ for \sptbounded,
while the retained state space decreases by $3.4\%$ and $3.3\%$,
respectively. The effect becomes increasingly pronounced as
$\hat{\varepsilon}$ grows.
At $\hat{\varepsilon}=0.40$, the peak retained state space decreases by
$31.8\%$ for \frtbounded and $34.2\%$ for \sptbounded, while solver time
decreases by $19.1\%$ and $27.9\%$, respectively. Activation cost decreases
by $17.7\%$ and $19.1\%$, while minimum achieved group coverage decreases
by about $17.5\%$ for both methods. At the same setting, the quota supplied
to the solver is approximately $17.4\%$ smaller than at
$\hat{\varepsilon}=0.01$.
Therefore,
$\hat{\varepsilon}$ provides a controlled computational--coverage
trade-off. Increasing it permits stronger state compression and reduces
solver time and activation cost, in exchange for a corresponding reduction
in required and achieved group coverage.

\subsection{Scalability}\label{sec:pokec}

We evaluate the scalable methods on the Pokec social
network~\cite{takac2012data}, using gender as the protected attribute.
The prepared connected graph contains 20,000 vertices and 262,846
undirected unit-cost edges. Nested connected graph instances are obtained
from deterministic breadth-first-search prefixes of the prepared graph.

We evaluate graph sizes of 1,000, 5,000, 10,000, and 20,000 vertices.
The corresponding original coverage requirements are 16, 32, 64, and 128
vertices per protected group. Each group contains twice as many candidate
vertices as required, so the original requirement always corresponds to
half of the available group members.

For each paired trial, the terminal is sampled from the current graph and
the same terminal and protected groups are supplied to every method.
We compare \sptbounded,
\frtbounded,
\pdmatched, and
\pdstrict.
The bounded FRT, bounded SPT, and matched primal--dual methods receive
the same per-instance relaxed coverage requirements, whereas \pdstrict uses
the original coverage targets.

\textbf{Results.}
\Cref{tab:pokec_results} reports mean runtime, activation cost, and achieved
coverage over the ten runs.
\begin{table*}[t]
	\centering\small
	\caption{Pokec scalability results over ten paired trials.
		Runtime and cost are mean $\pm$ standard deviation and coverage is the
		mean achieved Blue/Red count. \frtbounded, \sptbounded,
		and \pdmatched use matched relaxed coverage requirements; \pdstrict uses
		the original requirements.}
	\label{tab:pokec_results}
	
	\setlength{\tabcolsep}{5.5pt}
	\begin{tabular}{
			r
			l
			S[table-format=3.3] @{\,$\pm$\,} S[table-format=3.3]
			S[table-format=3.1] @{\,$\pm$\,} S[table-format=2.1]
			r
		}
		\toprule
		\textbf{Vertices} &
		\textbf{Method} &
		\multicolumn{2}{c}{\textbf{Runtime (s)}} &
		\multicolumn{2}{c}{\textbf{Cost}} &
		\textbf{Coverage B/R} \\
		\midrule
		
		$1{,}000$
		& \sptbounded
		& 0.041 & 0.022 & 44.2 & 2.9 & $16.0/16.0$ \\
		$1{,}000$
		& \frtbounded
		& 0.122 & 0.035 & 69.0 & 5.4 & $16.3/16.0$ \\
		$1{,}000$
		& \pdmatched
		& 0.437 & 0.336 & 49.4 & 3.5 & $16.0/16.2$ \\
		$1{,}000$
		& \pdstrict
		& 0.441 & 0.339 & 49.4 & 3.5 & $16.0/16.2$ \\
		
		\midrule
		
		$5{,}000$
		& \sptbounded
		& 0.269 & 0.053 & 101.1 & 6.4 & $31.0/31.0$ \\
		$5{,}000$
		& \frtbounded
		& 1.816 & 0.475 & 165.5 & 8.4 & $31.0/31.2$ \\
		$5{,}000$
		& \pdmatched
		& 14.02 & 19.52 & 101.6 & 5.0 & $31.2/31.0$ \\
		$5{,}000$
		& \pdstrict
		& 13.33 & 18.93 & 105.2 & 5.1 & $32.1/32.0$ \\
		
		\midrule
		
		$10{,}000$
		& \sptbounded
		& 1.482 & 0.255 & 194.1 & 8.1 & $61.0/61.0$ \\
		$10{,}000$
		& \frtbounded
		& 9.694 & 2.583 & 310.6 & 14.5 & $61.0/61.1$ \\
		$10{,}000$
		& \pdmatched
		& 127.14 & 109.62 & 195.5 & 4.9 & $61.1/61.4$ \\
		$10{,}000$
		& \pdstrict
		& 142.53 & 145.34 & 204.9 & 5.3 & $64.1/64.4$ \\
		
		\midrule
		
		$20{,}000$
		& \sptbounded
		& 13.58 & 3.78 & 385.8 & 12.3 & $122.0/122.0$ \\
		$20{,}000$
		& \frtbounded
		& 50.80 & 15.83 & 635.4 & 27.2 & $122.7/122.5$ \\
		$20{,}000$
		& \pdmatched
		& 454.39 & 484.36 & 381.8 & 11.1 & $122.0/122.1$ \\
		$20{,}000$
		& \pdstrict
		& 403.74 & 393.57 & 401.3 & 12.3 & $128.1/128.1$ \\
		
		\bottomrule
	\end{tabular}
\end{table*}
\sptbounded is faster than
\frtbounded at every graph size.
At 20,000 vertices, the respective mean runtimes are $13.58$~s and
$50.80$~s, a $3.74\times$ difference.
Because both methods receive the same coverage requirements, their costs can
also be compared directly. \sptbounded has lower cost in
all 40 paired trials; at 20,000 vertices the mean costs are $385.8$ and
$635.4$, respectively.
The runtime difference between
\sptbounded and \pdmatched becomes large as the
graph grows. At 20,000 vertices, their mean runtimes are $13.58$~s and
$454.39$~s, respectively, a $33.47\times$ ratio.
Their mean costs at the same size are instead very similar:
$385.8$ for \sptbounded and $381.8$ for the matched primal--dual method.

SPT is faster because the shortest-path-tree reduction removes most vertices before the dynamic
program is run. The mean retained tree contains 112.8, 247.8, 489.1, and
990.4 vertices across the four graph sizes, corresponding to pruning
$88.7\%$, $95.0\%$, $95.1\%$, and $95.0\%$ of the original shortest-path
tree.
The runtime at 20,000 vertices illustrates the resulting
difference. \frtbounded spends on average $25.69$~s
constructing the embedding, $11.26$~s in the tree dynamic program, and
$13.45$~s projecting the result back to the graph.
\sptbounded spends only $1.07$~s constructing and pruning
the direct tree and $12.48$~s in the dynamic program, with no embedding or
projection stage. The practical advantage on Pokec therefore comes mainly
from avoiding the FRT machinery and from reducing the graph to a small
colored core before optimization.

\section{Conclusion and future work}\label{sec:conclusion}

We introduced the \emph{Fair Minimum Labeling} problem for minimizing temporal activation cost under group-wise reachability constraints. We characterized the single-terminal case via rooted Covering Steiner, developed FRT- and SPT-based algorithms, and showed experimentally that enforcing group-wise coverage can improve balanced performance while SPT-FML scales effectively.

Future work includes results for multiple terminals, improved approximations, and extensions to  online, and distributed settings.

\clearpage

\appendix
\section*{Appendix}

\begin{algorithm2e}[htb]
	\caption{Bottom-up Label Computation on the Tree $T$}
	\label{alg:bottomup-tree-labels}
	\Input{Rooted tree $T = (V_T, E_T)$ with edge weights $w(u,v)$, coloring $\mathrm{col} : V_T \to \{B, R, \emptyset\}$.}
	\Output{At each node $v \in V_T$, a set $\mathrm{Labels}(v)$ of non-dominated labels $(b, r, \mathrm{cost})$.}

	\ForEach{node $v\in V_T$ in post-order}{
		$b_v \gets [\mathrm{col}(v) = B]$, \quad $r_v \gets [\mathrm{col}(v) = R]$\;
		$A \gets \{(b_v,\, r_v,\, 0)\}$ \tcp*{$v$ itself is always part of its own selected subtree}
		
		\tcc{For the bicriteria variant, replace the sequential child loop below by
			a balanced merge of the contribution sets $L^+_{v,u}$ from
			\Cref{def:mergedepth}, applying \textsc{Prune} after each pairwise merge.
			The exact variant may use the sequential loop below.}
		\ForEach{child $u$ of $v$}{
			$A' \gets \emptyset$\;
			\ForEach{$(b_1, r_1, \mathrm{cost}_1) \in A$}{
				Add $(b_1, r_1, \mathrm{cost}_1)$ to $A'$ \tcp*{do not select $u$'s subtree}
				\ForEach{$(b_2, r_2, \mathrm{cost}_2) \in \mathrm{Labels}(u)$}{
					Add $(b_1 + b_2,\, r_1 + r_2,\, \mathrm{cost}_1 + \mathrm{cost}_2 + w(v,u))$ to $A'$\;
				}
			}
			$A \gets \textsc{Prune}(A')$ \tcp*{exact variant: prune per $(b,r)$; bicriteria variant: use the balanced merge schedule above}%
		}
		$\mathrm{Labels}(v) \gets A$\;
	}
\end{algorithm2e}

\section{Omitted Proofs}\label{appendix}

\begin{observation}\label{obs:generalization}
	The FML problem generalizes both the Minimum Labeling (ML) and the Minimum Steiner Labeling (MSL) problems.
\end{observation}

\begin{proof}
	In ML, the goal is to make the entire graph temporally connected. We set $\terminals = V$, use a single color $c_1$ with $c(v)=\{c_1\}$ for all $v \in V$, put $\rho = |\terminals| = |V|$, and require $q_{c_1} = |V|$. Feasibility then says that all $|V|$ nodes reach all $|V|$ terminals, which is exactly all-pairs temporal connectivity, and the objective is the number of labels, as in ML.

	In MSL, the goal is to ensure temporal connectivity among a set $\terminals \subseteq V$. We use one color $c_1$ with $V_{c_1} = \terminals$, leave all other nodes uncolored, put $\rho = |\terminals|$ and require $q_{c_1} = |\terminals|$. Feasibility says that every terminal reaches every terminal, which is exactly the MSL condition, and again the objectives coincide.
\end{proof}

\begin{proof}[Proof of \Cref{theorem:maincomplexity}]
	\emph{Membership in NP.} We use \Cref{lemma:steiner}. If a single-terminal instance is a YES instance, then it has a feasible \qst subtree $S$ with $|E(S)|\le\min\{\budget,n-1\}$, and the labeling built in the ($\le$) direction of that lemma assigns one timestamp from $[n-1]$ to each edge of $S$ and nothing elsewhere. Such a labeling is a certificate of size $\mathcal{O}(n\log n)$, i.e., polynomial in the input length regardless of how the budget $\budget$ is encoded, and it is verified by a time-respecting BFS from every node followed by counting, for each color, the nodes of $V_c$ that reach $t$.

	For general FML, any YES-instance has a certificate with at most
	$2(n-1)$ labels: if the budget is at least $2(n-1)$, a temporally
	connected spanning-tree labeling with two labels per edge is feasible,
	while otherwise any feasible labeling already has fewer than
	$2(n-1)$ labels.

	\emph{Hardness.} We reduce from \textsc{Set Cover}. Let $(U,\mathcal{S})$ be an instance with universe $U=\{u_1,\dots,u_N\}$ and sets $\mathcal{S}=\{S_1,\dots,S_M\}$, every element lying in some set. Build the star
	\[ V = \{r\}\cup\{s_i : i\in[M]\}, \qquad E = \bigl\{\{r,s_i\} : i\in[M]\bigr\}, \]
	with terminal set $\terminals=\{r\}$ and $\rho=1$. Take one color per element, $\mathcal{C}=\{c_1,\dots,c_N\}$, colour each set node by the elements it contains,
	\[ c(s_i) = \{c_j : u_j\in S_i\}, \]
	leave $r$ uncolored, and set every requirement to $q_{c_j}=1$. The construction is linear in the size of the Set Cover instance, and $\numgroups = N$ exactly.

	Since $s_i$ is adjacent only to $r$, the node $s_i$ temporally reaches $r$ if and only if the edge $\{r,s_i\}$ carries at least one label, and no other node needs to be considered. Hence a labeling $\lambda$ is feasible if and only if the set $\mathcal{S}_\lambda = \{S_i : \lambda(\{r,s_i\})\neq\emptyset\}$ covers $U$, and an optimal $\lambda$ uses exactly one timestamp on each such edge and none elsewhere. Therefore
	\[ \mathrm{OPT}_{\mathrm{FML}} = \mathrm{OPT}_{\mathrm{SC}}(U,\mathcal{S}), \]
	and the map is an approximation-preserving reduction: from any labeling of cost $k$ we read off a cover of size at most $k$, and conversely. NP-hardness of the decision version follows from NP-hardness of Set Cover~\cite{karp1972}, and with the first part the decision version is NP-complete.

	\emph{Gap.} Because the reduction preserves the objective value exactly and $\numgroups=N$, Feige's theorem~\cite{feige1998threshold} transfers verbatim: for every fixed $\epsilon>0$, a polynomial-time $(1-\epsilon)\ln\numgroups$-approximation for single-terminal FML would give a $(1-\epsilon)\ln N$-approximation for Set Cover, which is impossible unless $\mathrm{NP}\subseteq\mathrm{DTIME}[N^{\mathcal{O}(\log\log N)}]$. Note that no conversion between $\ln N$ and $\ln|V|$ is involved, so the coefficient is not lost.
\end{proof}

\begin{proof}[Proof of \Cref{theorem:disjointcolors}]
	The construction above puts several colors on one node. To obtain the same bound with at most one color per node, we give every set-element incidence a private node of degree one, so that no colored node can serve as a relay for another, and we attach each set node to the terminal through a private chain whose length dominates the budget.

	Let $(U,\mathcal{S})$ be as above, fix a constant $\delta\in(0,1)$ and set $L := \lceil N/\delta\rceil$.
	\begin{itemize}
		\item A terminal $r$, with $\terminals=\{r\}$ and $\rho=1$.
		\item For each $S_i$ a node $s_i$ and a chain of exactly $L$ edges from $s_i$ to $r$, namely $s_i = w_{i,0}, w_{i,1},\dots,w_{i,L-1}, w_{i,L} = r$, where the $w_{i,\ell}$ for $1\le\ell\le L-1$ are fresh nodes private to $S_i$.
		\item For each incidence $u_j\in S_i$ a pendant node $v_{j,i}$ with the single edge $\{v_{j,i},s_i\}$, so $\deg(v_{j,i})=1$.
		\item Colors $\mathcal{C}=\{c_1,\dots,c_N\}$ with $c(v_{j,i})=\{c_j\}$ and $c(x)=\emptyset$ for every other node $x$; all requirements $q_{c_j}=1$.
		\item Budget $\budget := N + \kappa L$, where $\kappa$ is the target cover size.
	\end{itemize}
	Each set contributes $s_i$ together with $L-1$ internal chain nodes, i.e., $L$ private nodes, so
	\[ |V| \;=\; 1 + M L + \sum_{i}|S_i| \;=\; \mathrm{poly}(N,M), \]
	and every node carries at most one color, with $\numgroups=N$.

	\emph{(Cover $\Rightarrow$ labeling.)} Let $\mathcal{S}'$ be a cover with $|\mathcal{S}'|\le\kappa$. For every $S_i\in\mathcal{S}'$ label the $\ell$-th chain edge $\{w_{i,\ell-1},w_{i,\ell}\}$ with the timestamp $\ell+1$, $\ell=1,\dots,L$. For every element $u_j$ pick some $S_i\in\mathcal{S}'$ with $u_j\in S_i$ and label $\{v_{j,i},s_i\}$ with the timestamp $1$. The sequence $1,2,\dots,L+1$ is strictly increasing along $v_{j,i}, s_i, w_{i,1},\dots,r$, so the requirement of color $c_j$ is met, at cost $N+|\mathcal{S}'|L\le \budget$.

	\emph{(Labeling $\Rightarrow$ cover.)} Let $\lambda$ be feasible with $|\lambda|\le \budget$. For each $j$ some node of color $c_j$ reaches $r$, say $v_{j,i(j)}$; as it has degree one, its unique incident edge carries a label, and these $N$ edges are pairwise distinct. Any walk leaving $s_i$ towards $r$ must use $\{s_i,w_{i,1}\}$, because all other neighbours of $s_i$ are pendants of degree one; inductively, a time-respecting path from $v_{j,i}$ to $r$ traverses all $L$ edges of chain $i$. Let $P$ be the set of fully labeled chains. Chains are pairwise vertex-disjoint apart from $r$ and disjoint from the pendant edges, so $N+|P|L\le|\lambda|\le N+\kappa L$ and hence $|P|\le\kappa$. Finally $\{S_i : i\in P\}$ is a cover, since each $j$ is served through some $i(j)\in P$ and $v_{j,i(j)}$ exists only if $u_j\in S_{i(j)}$.

	The two directions give $\mathrm{OPT}_{\mathrm{FML}} = N + L\cdot\mathrm{OPT}_{\mathrm{SC}}$ exactly. For the gap, Feige's theorem makes it hard to distinguish $\mathrm{OPT}_{\mathrm{SC}}\le\kappa$ from $\mathrm{OPT}_{\mathrm{SC}} > (1-\delta)\ln N\cdot\kappa$. In the first case $\mathrm{OPT}_{\mathrm{FML}}\le N+\kappa L$; in the second $\mathrm{OPT}_{\mathrm{FML}} > L(1-\delta)\ln N\cdot\kappa$. Since $N\le\delta L\le\delta L\kappa$ we have $N+\kappa L\le(1+\delta)L\kappa$, so the ratio is at least $\frac{1-\delta}{1+\delta}\ln N = \frac{1-\delta}{1+\delta}\ln\numgroups$, and letting $\delta\to0$ gives the claim. Here the additive term $N$ is discharged because $N\le\delta L$, not merely because $L>N$.
\end{proof}

\begin{proof}[Proof of \Cref{theorem:runningtimetree}]
	Correctness follows from the weighted extension of \Cref{lemma:steiner}: an optimal $\mathrm{FML}_w$ solution on the tree is a minimum-weight subtree containing the root with the prescribed numbers of colored vertices, and one timestamp per selected edge suffices. The dynamic program computes exactly such a subtree: the label $(b,r,c)$ at $v$ records that the selected part of $v$'s subtree contains $b$ blue and $r$ red vertices at total edge weight $c$, and since the cost of the remainder of the solution depends on the child's contribution only through $(b,r)$, keeping one minimum-cost label per pair $(b,r)$ is lossless.

	For the running time, let $C = \mathcal{O}(n^2)$ bound the number of retained labels per node. Merging two label sets of size at most $C$ takes $\mathcal{O}(C^2) = \mathcal{O}(n^4)$ time: we form all pairs, add counts and costs plus the connecting edge weight, and keep one minimum-cost label per $(b,r)$, which restores the bound $C$. Each merge corresponds to one tree edge, and the contracted FRT tree has $\mathcal{O}(n)$ edges, since it has at most $2n-1$ nodes, so the total is $\mathcal{O}(n^5)$.

	Storing all retained merge tables together with standard backpointers uses $\mathcal{O}(n^3)$ space and permits reconstruction of the selected subtree.
\end{proof}

Before proving \Cref{theorem:weighted_approx} we show the following, stated in terms of the merge computation of \Cref{def:mergedepth}.

\begin{lemma}\label{lemma:weighted_approx_bound}
	Let $\varepsilon \in (0,1]$ and let $z$ be a merge-computation node with pruning depth $d_z$. For every exact label $(b,r,c)$ realizable at $z$ there is a label $(b',r',c')$ kept by the bucketing algorithm at $z$ with
	\begin{itemize}
		\item[(i)] $c' \le c$, and
		\item[(ii)] $b' \ge b/(1+\varepsilon)^{d_z}$ and $r' \ge r/(1+\varepsilon)^{d_z}$.
	\end{itemize}
\end{lemma}

\begin{proof}
	Induction on $d_z$. If $d_z=0$, no bucket-pruned merge has yet occurred, so every exact contribution state at $z$, including the skip state $(0,0,0)$, is present exactly; the claim therefore holds with equality.

	Let $d_z > 0$ with children $z_1, z_2$ of pruning depths at most $d_z - 1$. An exact label at $z$ arises by combining exact labels $(b_i, r_i, c_i)$ at $z_i$, giving $b = b_1 + b_2$, $r = r_1 + r_2$ and $c = c_1 + c_2$; the connecting edge costs are already included in the contribution sets $L^+_{v,u}$. By the induction hypothesis the algorithm keeps labels $(b'_i, r'_i, c'_i)$ at $z_i$ with $c'_i \le c_i$ and $b'_i \ge b_i/(1+\varepsilon)^{d_z - 1}$, and likewise for $r$. The algorithm therefore generates the candidate
	\[ b'' = b'_1 + b'_2 \ \ge\ \frac{b_1 + b_2}{(1+\varepsilon)^{d_z-1}} = \frac{b}{(1+\varepsilon)^{d_z-1}}, \qquad c'' = c'_1 + c'_2 \ \le\ c, \]
	and symmetrically for $r''$. The candidate falls into a bucket pair, from which the algorithm retains a label $(b',r',c')$ of minimum cost, so $c' \le c'' \le c$, which is (i). For (ii), $b'$ and $b''$ lie in the same bucket. If that bucket is bucket $0$ then $b'' = 0$, hence $b = 0$ and the claim is trivial; otherwise \Cref{def:buckets} gives $b' > b''/(1+\varepsilon)$, so
	\[ b' \ \ge\ \frac{b''}{1+\varepsilon} \ \ge\ \frac{b}{(1+\varepsilon)^{d_z}} , \]
	and symmetrically for $r'$. Adding a vertex's local color contribution exactly, without another pruning step, preserves these inequalities.
\end{proof}

\begin{proof}[Proof of \Cref{theorem:weighted_approx}]
	By \Cref{lemma:weighted_approx_bound}, the strict optimum has a retained representative $(\bar b,\bar r,\bar c)$ with $\bar c\le c_{\mathrm{opt}}$, $\bar b\ge b_{\mathrm{opt}}/\xi\ge q_B/\xi$, and $\bar r\ge r_{\mathrm{opt}}/\xi\ge q_R/\xi$. Since the counts are integral, $\bar b\ge\widetilde q_B$ and $\bar r\ge\widetilde q_R$. Hence the relaxed-feasible root set is nonempty, and the minimum-cost label $(b',r',c')$ selected from it satisfies $c'\le\bar c\le c_{\mathrm{opt}}$. The coverage bounds in (ii) hold directly from the relaxed root-state selection.

	For the resource bounds, every label set handled by the algorithm has size at most $\beta = \mathcal{O}(\varepsilon^{-2}\log^2 n)$ by \Cref{def:buckets}, so each merge costs $\mathcal{O}(\beta^2) = \mathcal{O}(\varepsilon^{-4}\log^4 n)$. The number of merges is $\sum_{v:\ell_v\ge2}(\ell_v - 1) < |E_T| = \mathcal{O}(n)$, giving the running time. There are $\mathcal{O}(n)$ merge nodes, so storing all retained tables and backpointers uses $\mathcal{O}(n\beta)=\mathcal{O}(n\varepsilon^{-2}\log^2 n)$ space.
\end{proof}

\begin{proof}[Proof of \Cref{cor:rescale}]
	If $D=0$, then $\xi=1$. For $D\ge1$, $\varepsilon=\hat\varepsilon/(2D)$ and hence $\xi = (1+\hat\varepsilon/(2D))^{D} \le e^{\hat\varepsilon/2} \le 1 + \hat\varepsilon$ for $\hat\varepsilon \le 1$. Substituting $\varepsilon=\hat\varepsilon/(2\max\{1,D\})$ into the running time of \Cref{theorem:weighted_approx} gives $\mathcal{O}(n\hat\varepsilon^{-4}\max\{1,D\}^4\log^4 n)$, and $D = \mathcal{O}(\log^2 n)$ by \Cref{def:mergedepth}.
\end{proof}

\begin{proof}[Proof of \Cref{theorem:main}]
	Fix the sampled tree $T$ and let $S$ be the subtree returned by the tree subroutine. By \Cref{lemma:projection} the algorithm outputs a labeling with
	\begin{equation}\label{eq:upper}
		|\lambda_G| \ \le\ \kappa\, C_T(S).
	\end{equation}

	For the lower bound, let $\lambda_G^*$ be an optimal FML solution in $G$ with $k_G^* = |\lambda_G^*|$. By \Cref{lemma:steiner} there is a subtree $S^*$ of $G$ with $t \in V(S^*)$, with $|E(S^*)| = k_G^*$, and containing at least $q_B$ blue and $q_R$ red vertices. Let $S_T^*$ be the minimal subtree of $T$ spanning the leaf set $V(S^*)$. Since the union of the tree paths between the endpoints of the edges of $S^*$ is connected and contains all of $V(S^*)$,
	\[ C_T(S^*_T) \ \le\ \sum_{\{u,v\} \in E(S^*)} d_T(u,v), \]
	and $S^*_T$ is feasible for the tree problem, because it contains $t$ and all colored leaves of $S^*$. Hence $C_{T,\mathrm{opt}} \le C_T(S^*_T)$. Both subroutines return $S$ with $C_T(S) \le C_{T,\mathrm{opt}}$ -- with equality for the exact one, and by \Cref{theorem:weighted_approx}(i) for the bicriteria one. Chaining with~\eqref{eq:upper},
	\[ |\lambda_G| \ \le\ \kappa\, C_T(S) \ \le\ \kappa\!\!\sum_{\{u,v\} \in E(S^*)}\!\! d_T(u,v). \]
	Taking expectations over $T \sim \mathcal{D}$ and using the expected stretch of \Cref{def:embedding} together with $d_G(u,v) = 1$ for every edge of $G$,
	\[ \mathbb{E}[|\lambda_G|] \ \le\ \kappa \!\!\sum_{\{u,v\} \in E(S^*)}\!\! \mathbb{E}[d_T(u,v)] \ \le\ \kappa\,\mathcal{O}(\log n) |E(S^*)| \ =\ \mathcal{O}(\log n)\, k_G^*. \]

	Feasibility of the output is immediate for the exact subroutine, and holds up to the factor $\xi$ for the bicriteria one by \Cref{theorem:weighted_approx}(ii); by the last sentence of \Cref{lemma:projection} the projection can only increase coverage.

	For the running time, computing $d_G$ takes $\mathcal{O}(nm)$ by breadth-first search from every vertex, and sampling the FRT tree takes $\mathcal{O}(n^2)$ additional time given the metric; the tree subroutine takes $\mathcal{O}(n^5)$ or $\mathcal{O}(n\varepsilon^{-4}\log^4 n)$ by \Cref{theorem:runningtimetree} and \Cref{theorem:weighted_approx}; the projection takes $\mathcal{O}(m + |E(\Pi(S))|) = \mathcal{O}(nm)$. The explicitly stored all-pairs metric uses $\mathcal{O}(n^2)$ space, giving the stated total space bounds.
\end{proof}

\section{Application Example}\label{app:example}
We give a concrete high-impact application scenario:
Consider a medical AI initiative deployed in a developing region to collect patient data for training a diagnostic model. The setting involves remote rural villages with limited connectivity and urban health hubs with better infrastructure. Devices operate under tight energy constraints and are duty-cycled, i.e., radios are only active during coordinated upload windows to conserve battery.

\begin{itemize}
	\item  \textbf{Nodes $V$:} Participants' smartphones, community health hubs, and the central server.

	\item  \textbf{Groups $B$, $R$:} For example, urban (near, well-connected) vs.~rural (remote) participants; we must connect an $\alpha$-fraction of each.
	Alternative groupings could capture other fairness dimensions, e.g., gender or ethnic communities (demographic fairness),
	users with high-end vs.~low-end devices (resource fairness),
	or solar-powered vs.~grid-powered nodes (energy fairness).

	\item  \textbf{Terminal $t$:} The central server.

	\item  \textbf{Edges $E$:} Potential short-range (Bluetooth/Wi-Fi Direct) links to hubs and long-range links from hubs to $t$.

	\item  \textbf{Temporal labeling $\lambda$:} Each activated edge is assigned a few global upload rounds, i.e., the only times radios are awake.

	\item  \textbf{Cost:} The total number of edge-time activations (each consumes energy when radios leave sleep mode).
\end{itemize}

\textbf{Why temporality matters.} Data must traverse time-respecting paths: if a rural phone $u$ forwards to hub $v$ in round $\tau_1$, then $v$ must forward towards $t$ in a later round $\tau_2>\tau_1$. Simply picking a static path is insufficient; we need an ordered activation plan consistent with duty-cycling. We note, by \Cref{lemma:steiner}, that with a single server the ordering constraint can always be met by one round per activated link; it becomes a real constraint when data must reach several servers, or when rounds are individually capacity-constrained.

\textbf{Outcome.} FML computes a sparse, globally scheduled activation plan with few edge-time activations that guarantees at least $\lceil\alpha|B|\rceil$ and $\lceil\alpha|R|\rceil$ nodes can reach $t$ temporally, while minimizing total wake-ups (edge-time labels). This aligns with real duty-cycled systems where communication is batch-scheduled to conserve energy and bandwidth.


\begin{thebibliography}{10}
	
	\bibitem{ahmadian2019clustering}
	Sara Ahmadian, Alessandro Epasto, Ravi Kumar, and Mohammad Mahdian.
	\newblock Clustering without over-representation.
	\newblock In {\em Proceedings of the 25th ACM SIGKDD International Conference
		on Knowledge Discovery \& Data Mining}, pages 267--275, 2019.
	
	\bibitem{akrida2017complexity}
	Eleni~C. Akrida, Leszek Gasieniec, George~B. Mertzios, and Paul~G. Spirakis.
	\newblock The complexity of optimal design of temporally connected graphs.
	\newblock {\em Theory of Computing Systems}, 61:907--944, 2017.
	
	\bibitem{axiotis2016size}
	Kyriakos Axiotis and Dimitris Fotakis.
	\newblock On the size and the approximability of minimum temporally connected
	subgraphs.
	\newblock In {\em Proceedings of the 43rd International Colloquium on Automata,
		Languages, and Programming}, pages 149:1--149:14, 2016.
	
	\bibitem{bartal1996probabilistic}
	Yair Bartal.
	\newblock Probabilistic approximation of metric spaces and its algorithmic
	applications.
	\newblock In {\em Proceedings of the 37th Conference on Foundations of Computer
		Science}, pages 184--193, 1996.
	
	\bibitem{bhore2022balanced}
	Sujoy Bhore, Sourav Chakraborty, Satyabrata Jana, Joseph S.~B. Mitchell,
	Supantha Pandit, and Sasanka Roy.
	\newblock The balanced connected subgraph problem.
	\newblock {\em Discrete Applied Mathematics}, 319:111--120, 2022.
	
	\bibitem{bose2019compositional}
	Avishek Bose and William Hamilton.
	\newblock Compositional fairness constraints for graph embeddings.
	\newblock In {\em Proceedings of the 36th International Conference on Machine
		Learning}, pages 715--724, 2019.
	
	\bibitem{byrka2013steiner}
	Jaros{\l}aw Byrka, Fabrizio Grandoni, Thomas Rothvo{\ss}, and Laura Sanit{\`a}.
	\newblock Steiner tree approximation via iterative randomized rounding.
	\newblock {\em Journal of the ACM}, 60(1):6:1--6:33, 2013.
	\newblock Preliminary version in STOC 2010.
	
	\bibitem{casteigts2012time}
	Arnaud Casteigts, Paola Flocchini, Walter Quattrociocchi, and Nicola Santoro.
	\newblock Time-varying graphs and dynamic networks.
	\newblock {\em International Journal of Parallel, Emergent and Distributed
		Systems}, 27(5):387--408, 2012.
	
	\bibitem{caton2024fairness}
	Simon Caton and Christian Haas.
	\newblock Fairness in machine learning: A survey.
	\newblock {\em ACM Computing Surveys}, 56(7), 2024.
	
	\bibitem{chen2024fairness}
	April Chen, Ryan~A. Rossi, Namyong Park, Puja Trivedi, Yu~Wang, Tong Yu,
	Sungchul Kim, Franck Dernoncourt, and Nesreen~K Ahmed.
	\newblock Fairness-aware graph neural networks: A survey.
	\newblock {\em ACM Transactions on Knowledge Discovery from Data}, 18(6):1--23,
	2024.
	
	\bibitem{chierichetti2017fair}
	Flavio Chierichetti, Ravi Kumar, Silvio Lattanzi, and Sergei Vassilvitskii.
	\newblock Fair clustering through fairlets.
	\newblock In {\em Advances in Neural Information Processing Systems},
	volume~30, 2017.
	
	\bibitem{chlebik2008steiner}
	Miroslav Chleb{\'i}k and Janka Chleb{\'i}kov{\'a}.
	\newblock The {S}teiner tree problem on graphs: Inapproximability results.
	\newblock {\em Theoretical Computer Science}, 406(3):207--214, 2008.
	
	\bibitem{dai2024comprehensive}
	Enyan Dai, Tianxiang Zhao, Huaisheng Zhu, Junjie Xu, Zhimeng Guo, Hui Liu,
	Jiliang Tang, and Suhang Wang.
	\newblock A comprehensive survey on trustworthy graph neural networks: Privacy,
	robustness, fairness, and explainability.
	\newblock {\em Machine Intelligence Research}, 21(6):1011--1061, 2024.
	
	\bibitem{dong2023fairness}
	Yushun Dong, Jing Ma, Song Wang, Chen Chen, and Jundong Li.
	\newblock Fairness in graph mining: A survey.
	\newblock {\em IEEE Transactions on Knowledge and Data Engineering},
	35(10):10583--10602, 2023.
	
	\bibitem{fakcharoenphol2003tight}
	Jittat Fakcharoenphol, Satish Rao, and Kunal Talwar.
	\newblock A tight bound on approximating arbitrary metrics by tree metrics.
	\newblock In {\em Proceedings of the 35th Annual ACM Symposium on Theory of
		Computing}, pages 448--455, 2003.
	
	\bibitem{feige1998threshold}
	Uriel Feige.
	\newblock A threshold of ln n for approximating set cover.
	\newblock {\em Journal of the ACM (JACM)}, 45(4):634--652, 1998.
	
	\bibitem{galimberti2020span}
	Edoardo Galimberti, Martino Ciaperoni, Alain Barrat, Francesco Bonchi, Ciro
	Cattuto, and Francesco Gullo.
	\newblock Span-core decomposition for temporal networks: Algorithms and
	applications.
	\newblock {\em ACM Transactions on Knowledge Discovery from Data (TKDD)},
	15(1):1--44, 2020.
	
	\bibitem{garg2005saving}
	Naveen Garg.
	\newblock Saving an epsilon: a 2-approximation for the $k$-{MST} problem in
	graphs.
	\newblock In {\em Proceedings of the 37th Annual ACM Symposium on Theory of
		Computing (STOC)}, pages 396--402, 2005.
	
	\bibitem{garg2000polylogarithmic}
	Naveen Garg, Goran Konjevod, and R.~Ravi.
	\newblock A polylogarithmic approximation algorithm for the group {S}teiner
	tree problem.
	\newblock {\em Journal of Algorithms}, 37(1):66--84, 2000.
	\newblock Preliminary version in SODA 1998, 253--259.
	
	\bibitem{gionis2024mining}
	Aristides Gionis, Lutz Oettershagen, and Ilie Sarpe.
	\newblock Mining temporal networks.
	\newblock In {\em Companion Proceedings of the ACM on Web Conference 2024},
	pages 1260--1263, 2024.
	
	\bibitem{gupta2006covering}
	Anupam Gupta and Aravind Srinivasan.
	\newblock An improved approximation ratio for the covering {S}teiner problem.
	\newblock {\em Theory of Computing}, 2(3):53--64, 2006.
	
	\bibitem{halperin2003polylogarithmic}
	Eran Halperin and Robert Krauthgamer.
	\newblock Polylogarithmic inapproximability.
	\newblock In {\em Proceedings of the 35th Annual ACM Symposium on Theory of
		Computing (STOC)}, pages 585--594, 2003.
	
	\bibitem{huaizhou2013fairness}
	SHI Huaizhou, R.~Venkatesha Prasad, Ertan Onur, and I.G.M.M. Niemegeers.
	\newblock Fairness in wireless networks: Issues, measures and challenges.
	\newblock {\em IEEE Communications Surveys \& Tutorials}, 16(1):5--24, 2013.
	
	\bibitem{jalali2023fairness}
	Zeinab~S. Jalali, Qilan Chen, Shwetha~M. Srikanta, Weixiang Wang, Myunghwan
	Kim, Hema Raghavan, and Sucheta Soundarajan.
	\newblock Fairness of information flow in social networks.
	\newblock {\em ACM Transactions on Knowledge Discovery from Data}, 17(6):1--26,
	2023.
	
	\bibitem{kairouz2021advances}
	Peter Kairouz, H.~Brendan McMahan, Brendan Avent, Aur{\'e}lien Bellet, Mehdi
	Bennis, Arjun~Nitin Bhagoji, Kallista Bonawitz, Zachary Charles, Graham
	Cormode, Rachel Cummings, et~al.
	\newblock Advances and open problems in federated learning.
	\newblock {\em Foundations and trends{\textregistered} in machine learning},
	14(1--2):1--210, 2021.
	
	\bibitem{karp1972}
	Richard~M. Karp.
	\newblock Reducibility among combinatorial problems.
	\newblock In {\em Proceedings of Complexity of Computer Computations}, pages
	85--103, 1972.
	
	\bibitem{kelly1997charging}
	Frank Kelly.
	\newblock Charging and rate control for elastic traffic.
	\newblock {\em European transactions on Telecommunications}, 8(1):33--37, 1997.
	
	\bibitem{kempe2000connectivity}
	David Kempe, Jon Kleinberg, and Amit Kumar.
	\newblock Connectivity and inference problems for temporal networks.
	\newblock In {\em Proceedings of the 32nd Annual ACM Symposium on Theory of
		Computing}, pages 504--513, 2000.
	
	\bibitem{kleindessner2019guarantees}
	Matth{\"a}us Kleindessner, Samira Samadi, Pranjal Awasthi, and Jamie
	Morgenstern.
	\newblock Guarantees for spectral clustering with fairness constraints.
	\newblock In {\em Proceedings of the 36th International Conference on Machine
		Learning}, pages 3458--3467, 2019.
	
	\bibitem{klobas2024complexity}
	Nina Klobas, George~B. Mertzios, Hendrik Molter, and Paul~G. Spirakis.
	\newblock The complexity of computing optimum labelings for temporal
	connectivity.
	\newblock {\em Journal of Computer and System Sciences}, 146:103564, 2024.
	
	\bibitem{konjevod2002covering}
	Goran Konjevod, R.~Ravi, and Aravind Srinivasan.
	\newblock Approximation algorithms for the covering {S}teiner problem.
	\newblock {\em Random Structures \& Algorithms}, 20(3):465--482, 2002.
	\newblock Preliminary version by Konjevod and Ravi in SODA 2000, 338--344.
	
	\bibitem{mao2024green}
	Yuyi Mao, Xianghao Yu, Kaibin Huang, Ying-Jun~Angela Zhang, and Jun Zhang.
	\newblock Green edge {AI}: A contemporary survey.
	\newblock {\em Proceedings of the IEEE}, 112(7):880--911, 2024.
	
	\bibitem{mertzios2013temporal}
	George~B. Mertzios, Othon Michail, Ioannis Chatzigiannakis, and Paul~G.
	Spirakis.
	\newblock Temporal network optimization subject to connectivity constraints.
	\newblock In {\em Proceedings of the 40th International Colloquium on Automata,
		Languages, and Programming}, pages 657--668, 2013.
	
	\bibitem{mertzios2019temporal}
	George~B. Mertzios, Othon Michail, and Paul~G. Spirakis.
	\newblock Temporal network optimization subject to connectivity constraints.
	\newblock {\em Algorithmica}, 81(4):1416--1449, 2019.
	
	\bibitem{michail2016introduction}
	Othon Michail.
	\newblock An introduction to temporal graphs: An algorithmic perspective.
	\newblock {\em Internet Mathematics}, 12(4):239--280, 2016.
	
	\bibitem{mutzel2019enumeration}
	Petra Mutzel and Lutz Oettershagen.
	\newblock On the enumeration of bicriteria temporal paths.
	\newblock In {\em International Conference on Theory and Applications of Models
		of Computation}, pages 518--535. Springer, 2019.
	
	\bibitem{nassef2022survey}
	Omar Nassef, Wenting Sun, Hakimeh Purmehdi, Mallik Tatipamula, and Toktam
	Mahmoodi.
	\newblock A survey: Distributed machine learning for {5G} and beyond.
	\newblock {\em Computer Networks}, 207:108820, 2022.
	
	\bibitem{navarro2024fair}
	Madeline Navarro, Samuel Rey, Andrei Buciulea, Antonio~G. Marques, and Santiago
	Segarra.
	\newblock Fair glasso: Estimating fair graphical models with unbiased
	statistical behavior.
	\newblock In {\em Advances in Neural Information Processing Systems},
	volume~37, pages 139589--139620, 2024.
	
	\bibitem{oettershagen2025edge}
	Lutz Oettershagen, Athanasios~L. Konstantinidis, and Giuseppe~F. Italiano.
	\newblock An edge-based decomposition framework for temporal networks.
	\newblock In {\em Proceedings of the Eighteenth ACM International Conference on
		Web Search and Data Mining}, pages 735--743, 2025.
	
	\bibitem{oettershagen2023higher}
	Lutz Oettershagen, Nils~M. Kriege, and Petra Mutzel.
	\newblock A higher-order temporal h-index for evolving networks.
	\newblock In {\em Proceedings of the 29th ACM SIGKDD Conference on Knowledge
		Discovery and Data Mining}, pages 1770--1782, 2023.
	
	\bibitem{oettershagen2026fair}
	Lutz Oettershagen and Othon Michail.
	\newblock Fair minimum labeling: Efficient temporal network activations for
	reachability and equity.
	\newblock {\em Advances in Neural Information Processing Systems},
	38:1087--1111, 2025.
	
	\bibitem{oettershagen2022temporal}
	Lutz Oettershagen, Petra Mutzel, and Nils~M. Kriege.
	\newblock Temporal walk centrality: ranking nodes in evolving networks.
	\newblock In {\em Proceedings of the ACM Web conference 2022}, pages
	1640--1650, 2022.
	
	\bibitem{oettershagen2024finding}
	Lutz Oettershagen, Honglian Wang, and Aristides Gionis.
	\newblock Finding densest subgraphs with edge-color constraints.
	\newblock In {\em Proceedings of the ACM Web Conference 2024}, pages 936--947,
	2024.
	
	\bibitem{rahman2019fairwalk}
	Tahleen Rahman, Bartlomiej Surma, Michael Backes, and Yang Zhang.
	\newblock Fairwalk: Towards fair graph embedding.
	\newblock In {\em Proceedings of the 28th International Joint Conference on
		Artificial Intelligence}, page 3289–3295, 2019.
	
	\bibitem{schwartz2020green}
	Roy Schwartz, Jesse Dodge, Noah~A. Smith, and Oren Etzioni.
	\newblock Green {AI}.
	\newblock {\em Communications of the ACM}, 63(12):54--63, 2020.
	
	\bibitem{takac2012data}
	Lubos Takac and Michal Zabovsky.
	\newblock Data analysis in public social networks.
	\newblock In {\em International scientific conference and international
		workshop present day trends of innovations}, volume~1, 2012.
	
	\bibitem{tsang2019group}
	Alan Tsang, Bryan Wilder, Eric Rice, Milind Tambe, and Yair Zick.
	\newblock Group-fairness in influence maximization.
	\newblock {\em arXiv preprint arXiv:1903.00967}, 2019.
	
	\bibitem{vaidya2000distributed}
	Nitin~H. Vaidya, Paramvir Bahl, and Seema Gupta.
	\newblock Distributed fair scheduling in a wireless {LAN}.
	\newblock In {\em Proceedings of the 6th Annual International Conference on
		Mobile Computing and Networking}, pages 167--178, 2000.
	
	\bibitem{vatter2023evolution}
	Jana Vatter, Ruben Mayer, and Hans-Arno Jacobsen.
	\newblock The evolution of distributed systems for graph neural networks and
	their origin in graph processing and deep learning: A survey.
	\newblock {\em ACM Computing Surveys}, 56(1):1--37, 2023.
	
	\bibitem{verbraeken2020survey}
	Joost Verbraeken, Matthijs Wolting, Jonathan Katzy, Jeroen Kloppenburg, Tim
	Verbelen, and Jan~S. Rellermeyer.
	\newblock A survey on distributed machine learning.
	\newblock {\em ACM Computing Surveys}, 53(2):1--33, 2020.
	
	\bibitem{windham2024fast}
	Dennis~Robert Windham, Caroline~J. Wendt, Alex Crane, Madelyn~J. Warr, Freda
	Shi, Sorelle~A. Friedler, Blair~D. Sullivan, and Aaron Clauset.
	\newblock Fast algorithms to improve fair information access in networks.
	\newblock {\em arXiv preprint arXiv:2409.03127}, 2024.
	
	\bibitem{zhou2024fairness}
	Zhuoping Zhou, Davoud Ataee~Tarzanagh, Bojian Hou, Qi~Long, and Li~Shen.
	\newblock Fairness-aware estimation of graphical models.
	\newblock In {\em Advances in Neural Information Processing Systems},
	volume~37, pages 17870--17909, 2024.
	
\end{thebibliography}
\end{document}